\documentclass[11pt]{article}
\usepackage{epsfig}
\usepackage{amsmath,amsfonts}

\newcommand{\sect}[1]{ \section{#1} \setcounter{equation}{0} }

\newcommand{\pslash}{p \! \! \! /}

\newcommand{\Nf}{N_{\!f}}

\newcommand{\MSbar}{\overline{\mbox{MS}}}

\begin{document}
\date{}
\title{\textbf{A study of the gauge invariant, nonlocal mass operator $\mathbf{Tr \int d^4x F_{\mu\nu}
(D^2)^{-1} F_{\mu\nu} }$ in Yang-Mills theories}}
\author{ \textbf{M.A.L. Capri$^a$\thanks{marcio@dft.if.uerj.br}} \ , \textbf{D. Dudal}$^{b}$\thanks{david.dudal@ugent.be}{\
} \ , \textbf{J.A. Gracey$^{c}$\thanks{jag@amtp.liv.ac.uk}} \ ,
\textbf{V.E.R. Lemes$^{a}$\thanks{vitor@dft.if.uerj.br}}  \ , \\\textbf{R.F. Sobreiro}$^{a}$\thanks{%
sobreiro@uerj.br}  \ , \textbf{S.P. Sorella}$^{a}$\thanks{%
sorella@uerj.br}{\ }{\ }\footnote{Work supported by FAPERJ, Funda{\c
c}{\~a}o de Amparo {\`a} Pesquisa do Estado do Rio de Janeiro, under
the program {\it Cientista do Nosso Estado}, E-26/151.947/2004.}  \
,  \textbf{H. Verschelde}$^{b}$\thanks{henri.verschelde@ugent.be} \\\\
\textit{$^{a}$\small{Departamento de F\'{\i }sica Te\'{o}rica}}\\
\textit{\small{Instituto de F\'{\i }sica, UERJ, Universidade do Estado do Rio de Janeiro}} \\
\textit{\small{Rua S{\~a}o Francisco Xavier 524, 20550-013 Maracan{\~a}}} \\
\textit{\small{Rio de Janeiro, Brasil}} \\[3mm]
\textit{$^{b}$\small{Ghent University}} \\
\textit{\small{Department of Mathematical Physics and Astronomy}} \\
\textit{\small{Krijgslaan 281-S9, B-9000 Gent, Belgium}}\\
[3mm] \textit{$^c$\small{Theoretical Physics Division} }\\
\textit{\small{Department of Mathematical Sciences}}\\
\textit{\small{University of Liverpool}}\\
\textit{\small{P.O. Box 147, Liverpool, L69 3BX, United Kingdom}} }
\maketitle

\begin{abstract}
\noindent The nonlocal mass operator $\mathrm{Tr} \int d^4x
F_{\mu\nu} (D^2)^{-1} F_{\mu\nu} $ is considered in Yang-Mills
theories in Euclidean space-time. It is shown that the operator
$\mathrm{Tr}\int d^4x F_{\mu\nu} (D^2)^{-1} F_{\mu\nu} $ can be
cast in local form through the introduction of a set of additional
fields. A local and polynomial action is thus identified. Its
multiplicative renormalizability is proven by means of the
algebraic renormalization in the class of linear covariant gauges.
The anomalous dimensions of the fields and of the mass operator
are computed at one loop order. A few remarks on the possible role
of this operator for the issue of the gauge invariance of the
dimension two condensates are outlined.
\end{abstract}
\vspace{-18cm} \hfill LTH--659 \vspace{16cm}
\newpage

\newpage



\sect{Introduction.} Dimension two condensates have received great
attention in recent years. These condensates might play an important
role for the infrared dynamics of Euclidean Yang-Mills theories, as
supported by the considerable amount of results obtained through
theoretical and phenomenological studies as well as from lattice
simulations \cite
{Cornwall:1981zr,Greensite:1985vq,Stingl:1985hx,Lavelle:1988eg,Gubarev:2000nz,Gubarev:2000eu,
Verschelde:2001ia,Kondo:2001nq, Kondo:2001tm,Dudal:2003vv,
Browne:2003uv,Dudal:2003gu, Dudal:2003by,
Dudal:2004rx,Browne:2004mk, Gracey:2004bk,
Li:2004te,Boucaud:2001st,Boucaud:2002nc,
Boucaud:2005rm, RuizArriola:2004en,Suzuki:2004dw,Gubarev:2005it,Furui:2005bu,Boucaud:2005xn,Chernodub:2005gz}%
. \newline
\newline
For instance, the gluon condensate $\left\langle A_{\mu }^{a}A_{\mu
}^{a}\right\rangle $ has been largely investigated in the Landau gauge. As
pointed out in \cite{Lavelle:1988eg}, this condensate enters the operator
product expansion (OPE) of the gluon propagator. Moreover, a combined OPE
and lattice analysis has shown that this condensate can account for the $%
1/Q^{2}$ corrections which have been reported \cite
{Boucaud:2001st,Boucaud:2002nc,Boucaud:2005rm,RuizArriola:2004en,Furui:2005bu,Boucaud:2005xn}
in the running of the coupling constant and in the gluon
correlation functions. An effective potential for $\left\langle
A_{\mu }^{a}A_{\mu }^{a}\right\rangle $ has been obtained and
evaluated in analytic form at two loop in \cite
{Verschelde:2001ia,Dudal:2003vv,Browne:2003uv,Browne:2004mk,Gracey:2004bk},
showing that a nonvanishing value of $\left\langle A_{\mu
}^{a}A_{\mu }^{a}\right\rangle $ is favoured as it lowers the
vacuum energy. As a consequence, a dynamical gluon mass is
generated. We also recall that, in the Landau gauge, the operator
$A_{\mu }^{a}A_{\mu }^{a}$ is $BRST$ invariant on-shell, a
property which has allowed for an all orders proof of its
multiplicative renormalizability. Its anomalous dimension is not
an independent parameter, being expressed as a combination of the
gauge $\beta - $function and of the anomalous dimension, $\gamma
_{A}$, of the gauge field $A_{\mu }^{a}$ \cite{Dudal:2002pq}. This
relation was conjectured and explicitly verified up to three-loop
order in \cite{Gracey:2002yt}.
\newline
\newline
The dimension two operator $A_{\mu }^{a}A_{\mu }^{a}$ has been proven to be
multiplicatively renormalizable to all orders in the more general class of
linear covariant gauges \cite{Dudal:2003np}. An effective potential for the
condensate $\left\langle A_{\mu }^{a}A_{\mu }^{a}\right\rangle $ in linear
covariant gauges has been evaluated in \cite{Dudal:2003by}, providing
evidence for a nonvanishing value $\left\langle A_{\mu }^{a}A_{\mu
}^{a}\right\rangle $ in these gauges. \newline
\newline
A renormalizable mass dimension two operator can be introduced in
other covariant renormalizable gauges, such as the Curci-Ferrari
and the maximal Abelian gauge.\ In the Curci-Ferrari gauge the
generalized gluon-ghost
operator $\left( \frac{1}{2}A_{\mu }^{a}A_{\mu }^{a}+\alpha \overline{c}%
^{a}c^{a}\right) $ is $BRST$ invariant on-shell, displaying
multiplicative
renormalizability to all orders \cite{Dudal:2003pe}. The fields $%
\overline{c}^{a}$, $c^{a}$ stand for the Faddeev-Popov ghosts, while $\alpha
$ denotes the gauge parameter. Evidence for a nonvanishing condensate $%
\left\langle \frac{1}{2}A_{\mu }^{a}A_{\mu }^{a}+\alpha \overline{c}%
^{a}c^{a}\right\rangle $ have been provided in \cite{Dudal:2003gu}. Note
that in the limit $\alpha \rightarrow 0$, corresponding to the Landau gauge,
the operator $\left( \frac{1}{2}A_{\mu }^{a}A_{\mu }^{a}+\alpha \overline{c}%
^{a}c^{a}\right) $ reduces to $A_{\mu }^{a}A_{\mu }^{a}$. A mixed
gluon-ghost operator, namely $\left( \frac{1}{2}A_{\mu }^{A}A_{\mu
}^{A}+\alpha \overline{c}^{A}c^{A}\right) $, can be introduced
also in the maximal Abelian gauge
\cite{Kondo:2001nq,Kondo:2001tm,Dudal:2004rx}. Here the color
index $A$ runs over the $N(N-1)$ off-diagonal generators of the
gauge group $SU(N)$, $A=1,....,N(N-1)$. As in the case of the
Curci-Ferrari gauge, this operator is $BRST$  invariant on-shell,
being multiplicatively renormalizable to all orders \cite
{Kondo:2001nq,Kondo:2001tm,Dudal:2004rx,Dudal:2003pe,Gracey:2005vu}.
Analytic evidence for a nonvanishing condensate $\left\langle \frac{1}{2}%
A_{\mu }^{A}A_{\mu }^{A}+\alpha \overline{c}^{A}c^{A}\right\rangle
$ in the maximal Abelian gauge can be found in
\cite{Dudal:2004rx}. We underline that a nonvanishing condensate
$\left\langle \frac{1}{2}A_{\mu }^{A}A_{\mu }^{A}+\alpha
\overline{c}^{A}c^{A}\right\rangle $ gives rise to the dynamical
mass generation for off-diagonal gluons, a result of great
relevance for the so-called Abelian dominance, supporting the dual
superconductivity picture for color confinement. An off-diagonal
gluon mass has also been reported in lattice simulations \cite
{Amemiya:1998jz,Bornyakov:2003ee}. \newline
\newline
Studies of the influence of these condensates on the gluon and ghost
propagators when the nonperturbative effects of the Gribov copies are taken
into account can be found in \cite
{Sobreiro:2004us,Dudal:2005na,Sobreiro:2005vn,Capri:2005tj}. The output of
these analysis is an infrared suppression of the components of the gluon
propagator in the aforementioned gauges, a feature in agreement with the
results available from lattice and Schwinger-Dyson studies \cite
{Marenzoni:1994ap,Leinweber:1998uu,Bonnet:2001uh,Langfeld:2001cz,
Cucchieri:2003di,Bloch:2003sk,Furui:2003jr,Silva:2004bv,Amemiya:1998jz,Bornyakov:2003ee,Giusti:1996kf, Giusti:1999im,Giusti:2000yc,vonSmekal:1997iss,vonSmekal:1997is,Atkinson:1997tu,Alkofer:2000wg,Watson:2001yv, Alkofer:2003jr}%
. \newline
\newline
Certainly, many aspects related to the dimension two condensates
deserve a better understanding. This is the case, for example, of
the gauge invariance, a central issue in order to give a precise
physical meaning to these condensates. A recent study of this
topic has been given in \cite
{Slavnov:2004rz,Slavnov:2005av,Bykov:2005tx}, where a set of
conditions which should ensure the independence of the condensate
$\left\langle A_{\mu }^{a}A_{\mu }^{a}\right\rangle $ from the
gauge parameter in the class of linear covariant gauges has been
proposed. \newline
\newline
In this work we pursue the study on the aspects of the gauge invariance of
the dimension two condensates. Our aim here is that of discussing the
possibility of introducing a suitable colorless dimension two operator $%
\mathcal{O}(A)$ which preserves gauge invariance
\begin{eqnarray}
\delta \mathcal{O}(A) &=&0\;,  \nonumber \\
\delta A_{\mu }^{a} &=&-D_{\mu }^{ab}\omega ^{b}\;,  \label{gauge0}
\end{eqnarray}
where $D_{\mu }^{ab}$ is the covariant derivative
\begin{equation}
D_{\mu }^{ab}=\delta ^{ab}\partial _{\mu }-gf^{abc}A_{\mu }^{c}\;.
\label{cov0}
\end{equation}
This is a difficult task, due to the lack of a local gauge
invariant mass term built up with gauge fields only. This problem
could be overcome by looking at nonlocal operators. However, even
if we allow for nonlocal operators, we cannot give up of the
requirement that a consistent computational framework, allowing to
carry out higher loop calculations, has to be at our disposal.
This is a strong requirement which, in practice, deeply constrains
the type of nonlocality allowed for the dimension two operator. As
a suitable proposition in order to obtain such a consistent
framework, we could demand that
the action to which the nonlocal gauge invariant operator $\mathcal{O}%
(A)$ is coupled, should have the property of being made local by the
introduction of a suitable set of additional fields.
\begin{itemize}
\item[(I.)] Therefore, denoting by $S_{\mathcal{O}}$ the term which accounts for
the introduction in the Yang-Mills action, $S_{YM}$, of the operator $%
\mathcal{O}(A)$ in its localized form, we require that
$S_{\mathcal{O}}$ is gauge invariant. \item[(II.)]  Also, on
physical grounds, we demand that the introduction of the operator
$\mathcal{O}(A)$ makes it possible to identify a quantized action
which is multiplicatively renormalizable, a feature which should
not be related to a specific choice of the gauge fixing term
$S_{gf}$, of course on the condition that the usual Yang-Mills
action $S$, quantized using the gauge fixing $S_{gf}$, thus
$S=S_{YM}+S_{gf}$, is renormalizable.
\end{itemize}
\noindent As we shall see, these conditions will lead us to consider
the nonlocal gauge invariant operator of mass dimension two
\begin{equation}
\mathcal{O}(A)=-\frac{1}{2}\int d^{4}xF_{\mu \nu }^{a}\left[ \left(
D^{2}\right) ^{-1}\right] ^{ab}F_{\mu \nu }^{b}\;.  \label{gm}
\end{equation}
Expression (\ref{gm}) can be made local by the introduction of a set of
additional fields. Moreover, we will be able to prove that it is possible to
identify a local and polynomial action which turns out to be
multiplicatively renormalizable to all orders. \newline
\newline
The identification of this action and the algebraic proof of its
renormalizability, as explicitly checked through the evaluation of
the one loop anomalous dimensions, are the main results of the
present investigation, signaling that the operator (\ref{gm})
could be relevant for a better understanding of the issue of the
gauge invariance of the dimension two gluon condensate. \newline
\newline
Besides the renormalizability, we should also provide a suitable framework
to discuss the possible condensation of the operator (\ref{gm}), \textit{%
i.e. }$\left\langle \mathcal{O}(A)\right\rangle \neq 0$, which would give
rise to the dynamical gluon mass generation. Although being out of the aim
of the present work, we remark that, in the Landau gauge, $\partial _{\mu
}A_{\mu }^{a}=0$, expression (\ref{gm}) reduces, to the first order, to the
mass operator $\int d^{4}xA_{\mu }^{a}A_{\mu }^{a}$,
\begin{equation}
-\frac{1}{2}\int d^{4}xF_{\mu \nu }^{a}\left[ \left( D^{2}\right)
^{-1}\right] ^{ab}F_{\mu \nu }^{b}=\int d^{4}xA_{\mu }^{a}A_{\mu }^{a}\;+\;%
\mathrm{higher\;order\;terms}\;.  \label{higho}
\end{equation}
Thus, it is not inconceivable that a nonvanishing condensate
$\left\langle A_{\mu }^{a}A_{\mu }^{a}\right\rangle $ $\neq 0$
might provide a support in favor of a nonvanishing condensation of
the
operator (\ref{gm}), \textit{i.e. }$\left\langle F\frac{1}{D^{2}}%
F\right\rangle \neq 0$. \newline
\newline
The plan of the work is as follows. In section 2 we give an
account of a set of nonlocal and gauge invariant mass operators
which can be introduced
in the Abelian case. These include the Abelian version of the operator $%
\mathcal{O}(A)$ of eq.(\ref{gm}), the operator $A_{\min }^{2}$ recently
discussed in \cite{Gubarev:2000nz,Gubarev:2000eu}, the Stueckelberg term as
well as the nonlocal mass operator $\int d^{4}xA_{\mu }^{T}A_{\mu }^{T}$,
where $A_{\mu }^{T}$ stands for the transverse component of the gauge field $%
A_{\mu }$, $A_{\mu }^{T}=$ $\left( \delta _{\mu \nu
}-\frac{\partial _{\mu }\partial _{\nu }}{\partial ^{2}}\right)
A_{\nu }$. Interestingly, in the Abelian case, it turns out that
all these gauge invariant operators can be proven to be
classically equivalent, \textit{i.e. }they reduce to the same
expression when the classical equations of motion are used. In
section 3 we present a detailed discussion of the non-Abelian
generalization of these mass operators. We shall see that all
operators introduced in the Abelian case possess a non-Abelian
gauge invariant extension. However, the classical equivalence
between them is now no longer valid. In particular, we point out
that, in the non-Abelian case, the mass operator of eq.(\ref{gm})
exhibits differences with respect to the operator $A_{\min }^{2}$.
As we shall see, the latter can be expressed as an infinite sum of
nonlocal terms, a feature which makes almost hopeless the
possibility of achieving a consistent localization procedure for a
generic choice of the gauge fixing condition. Section 4 is devoted
to the study of the localization procedure of the mass operator
(\ref{gm}) and of the rich symmetry content of the resulting
action. In section 5, the identification of a suitable local and
polynomial action is provided. Its multiplicative
renormalizability in the class of covariant linear gauges will be
established by means of the algebraic renormalization. Having
developed the general properties of the mass operator, we devote
section 6 to the computation of its anomalous dimension at one
loop. Our conclusions are presented in section 7. For the benefit
of the reader, we have found useful to collect in several
Appendices the explicit derivation of some relevant features of
the various mass operators considered in this work.

\sect{Mass operators in the Abelian case.} In this section we shall
discuss a set of nonlocal gauge invariant mass operators which can
be added to the Maxwell action
\begin{equation}
\frac{1}{4}\int d^{4}xF_{\mu \nu }F_{\mu \nu }\;.  \label{max0}
\end{equation}
Perhaps, the simplest way of introducing a gauge invariant mass term is
through the nonlocal gauge invariant variable $A_{\mu }^{T}$
\begin{equation}
A_{\mu }^{T}=\left( \delta _{\mu \nu }-\frac{\partial _{\mu }\partial _{\nu }%
}{\partial ^{2}}\right) A_{\nu }\;.  \label{trans0}
\end{equation}
Expression (\ref{trans0}) is recognized to be the transverse component of
the gauge field, $\partial _{\mu }A_{\mu }^{T}=0$, and is invariant under
the gauge transformations, \textit{i.e. }
\begin{equation}
\delta A_{\mu }^{T}=0\;,  \label{inv0}
\end{equation}
with
\begin{equation}
\delta A_{\mu }=-\partial _{\mu }\omega \;,  \label{transf0}
\end{equation}
Thus, for the gauge invariant mass term one writes
\begin{equation}
\mathcal{O}_{1}(A)=\int d^{4}xA_{\mu }^{T}A_{\mu }^{T}\;.  \label{op1}
\end{equation}
A second possibility of introducing an invariant mass term is provided by
the operator $A_{\min }^{2}$, which has been recently analysed in \cite
{Gubarev:2000nz,Gubarev:2000eu}. The operator $A_{\min }^{2}$ is obtained by
minimizing the quantity $\int d^{4}xA_{\mu }A_{\mu }$ with respect to the
gauge transformations, namely
\begin{equation}
\mathcal{O}_{2}(A)=A_{\min }^{2}=\mathrm{min}\int d^{4}xA_{\mu }A_{\mu }\;.
\label{op2}
\end{equation}
Making use of the decomposition of the gauge field $A_{\mu }$ into
transverse and longitudinal parts
\begin{eqnarray}
A_{\mu } &=&A_{\mu }^{T}+A_{\mu }^{L}\;,  \nonumber \\
A_{\mu }^{T} &=&\left( \delta _{\mu \nu }-\frac{\partial _{\mu }\partial
_{\nu }}{\partial ^{2}}\right) A_{\nu }\;,  \nonumber \\
A_{\mu }^{L} &=&\frac{\partial _{\mu }\partial _{\nu }}{\partial ^{2}}A_{\nu
}\;,  \label{decomp0}
\end{eqnarray}
it follows that
\begin{equation}
\int d^{4}xA_{\mu }A_{\mu }=\int d^{4}xA_{\mu }^{T}A_{\mu }^{T}+\int
d^{4}xA_{\mu }^{L}A_{\mu }^{L}\;.  \label{decomp1}
\end{equation}
Observe that both terms in equation (\ref{decomp1}) are positive definite.
Moreover, as discussed in \cite{Gubarev:2000nz,Gubarev:2000eu}, the
functional $\int d^{4}xA_{\mu }A_{\mu }$ achieves its minimum when $\partial
_{\mu }A_{\mu }=0$, \textit{i.e.} $A_{\mu }^{L}=0$, so that
\begin{equation}
\mathcal{O}_{1}(A)=\mathcal{O}_{2}(A)\;,  \label{equiv0}
\end{equation}
which establishes the equivalence between expressions (\ref{op1}) and (\ref
{op2}). It is worth mentioning that the gauge invariant functional $A_{\min
}^{2}$ has been proven to be an order parameter for the study of the phase
transition of compact three-dimensional QED \cite{Gubarev:2000eu}%
. \newline
\newline
A third possibility of introducing an invariant mass operator in the Abelian
case is by means of the Stueckelberg term \cite{Ruegg:2003ps}
\begin{equation}
\mathcal{O}_{3}(A)=\int d^{4}x\left( A_{\mu }+\partial _{\mu }\phi \right)
^{2}\;,  \label{op3}
\end{equation}
where $\phi $ is a dimensionless scalar field. Expression (\ref{op3}) is
left invariant by the following transformations
\begin{eqnarray}
\delta A_{\mu } &=&-\partial _{\mu }\omega \;,  \nonumber \\
\delta \phi &=&\omega \;.  \label{gauge1}
\end{eqnarray}
The mass term (\ref{op3}) can be rewritten in the form of a $U(1)$ gauged $%
\sigma $-model, by introducing the variable
\begin{equation}
U=e^{ie\phi }\;.  \label{quant0}
\end{equation}
Thus
\begin{equation}
\mathcal{O}_{3}(A)=\int d^{4}x\left( A_{\mu }-\frac{i}{e}U^{-1}\partial
_{\mu }U\right) ^{2}\;.  \label{op3a}
\end{equation}
Transformations (\ref{gauge1}) read now
\begin{eqnarray}
A_{\mu } &\rightarrow &A_{\mu }+\frac{i}{e}V^{-1}\partial _{\mu }V\;,
\nonumber \\
U &\rightarrow &UV\;,  \label{gauge2}
\end{eqnarray}
with
\begin{equation}
V=e^{ie\omega }\;.  \label{quant1}
\end{equation}
One checks that the quantity $\left( A_{\mu
}-\frac{i}{e}U^{-1}\partial _{\mu }U\right) $ is left invariant by
the transformations (\ref{gauge2}). Analogously to the operator
$\mathcal{O}_{2}(A)$, expression (\ref{op3}) can be proven to be
classically equivalent to the mass term of equation (\ref {op1}).
This is easily seen by looking at the equations of motion which
follow from the gauge invariant action
\begin{equation}
S=\frac{1}{4}\int d^{4}xF_{\mu \nu }F_{\mu \nu }+\frac{m^{2}}{2}\int
d^{4}x\left( A_{\mu }+\partial _{\mu }\phi \right) ^{2}\;,
\end{equation}
namely
\begin{eqnarray}
\partial _{\nu }F_{\mu \nu }+m^{2}\left( A_{\mu }+\partial _{\mu }\phi
\right) &=&0\;,  \nonumber \\
\partial ^{2}\phi +\partial _{\mu }A_{\mu } &=&0\;.  \label{eq0}
\end{eqnarray}
In particular, from the second equation of (\ref{eq0}), we obtain
\begin{equation}
\phi =-\frac{1}{\partial ^{2}}\partial A\;,  \label{eq1}
\end{equation}
so that
\begin{equation}
\mathcal{O}_{3}(A)=\int d^{4}x\left( A_{\mu }-\frac{\partial _{\mu }\partial
_{\nu }}{\partial ^{2}}A_{\nu }\right) ^{2}=\int d^{4}xA_{\mu }^{T}A_{\mu
}^{T}\;.
\end{equation}
Thus
\begin{equation}
\mathcal{O}_{3}(A)=\mathcal{O}_{1}(A)\;,  \label{equiv1}
\end{equation}
which establishes the classical equivalence between expressions (\ref{op1})
and (\ref{op3}). Also, from (\ref{eq1}) one sees that the scalar field $\phi
$ is related to the longitudinal mode of the gauge field $A_{\mu }$. \newline
\newline
Finally, a fourth mass operator can be introduced by considering the
nonlocal quantity
\begin{equation}
\mathcal{O}_{4}(A)=-\frac{1}{2}\int d^{4}xF_{\mu \nu }\frac{1}{\partial ^{2}}%
F_{\mu \nu }\;.  \label{op4}
\end{equation}
Again, this term is seen to be equivalent to expression (\ref{op1}). In fact
\begin{eqnarray}
\mathcal{O}_{4}(A) &=&-\frac{1}{2}\int d^{4}x\left( \partial _{\mu }A_{\nu
}-\partial _{\nu }A_{\mu }\right) \frac{1}{\partial ^{2}}\left( \partial
_{\mu }A_{\nu }-\partial _{\nu }A_{\mu }\right)  \nonumber \\
&=&\frac{1}{2}\int d^{4}x\left[ A_{\nu }\frac{1}{\partial ^{2}}\left(
\partial ^{2}A_{\nu }-\partial _{\mu }\partial _{\nu }A_{\mu }\right)
+A_{\mu }\frac{1}{\partial ^{2}}\left( \partial ^{2}A_{\mu }-\partial _{\nu
}\partial _{\mu }A_{\nu }\right) \right]  \nonumber \\
&=&\int d^{4}xA_{\nu }\left( A_{\nu }-\frac{\partial _{\mu }\partial _{\nu }%
}{\partial ^{2}}A_{\mu }\right) =\int d^{4}xA_{\mu }^{T}A_{\mu }^{T}\;,
\label{eqq}
\end{eqnarray}
thus
\begin{equation}
\mathcal{O}_{4}(A)=\mathcal{O}_{1}(A)\;.  \label{equiv2}
\end{equation}
Albeit nonlocal, the operator $\left( \ref{op4}\right) $ can be made local
through the introduction of suitable additional fields. More precisely, in
the present case, one has
\begin{equation}
-\frac{1}{4}m^{2}\int d^{4}xF_{\mu \nu }\frac{1}{\partial ^{2}}F_{\mu \nu
}\rightarrow \int d^{4}x\left( \frac{1}{4}\overline{B}_{\mu \nu }\partial
^{2}B_{\mu \nu }+\frac{im}{4}\left( F_{\mu \nu }B_{\mu \nu }-F_{\mu \nu }%
\overline{B}_{\mu \nu }\right) \right) \;,  \label{g5}
\end{equation}
where $\overline{B}_{\mu \nu }$ and $B_{\mu \nu }$ are a pair of
antisymmetric complex fields and $m$ is a mass parameter. Eliminating $%
\overline{B}_{\mu \nu }$ and $B_{\mu \nu }$ by means of their equations of
motion, one gets back the nonlocal action $\left( \ref{op4}\right) $. One
sees thus that, once cast in the local form, expression $\left( \ref{op4}%
\right) $ looks renormalizable by power counting. It turns out in fact that,
in the Abelian case, the localized term in the right hand side of eq.$\left(
\ref{g5}\right) $ can be added to the usual $QED$\ Lagrangian without
destroying its renormalizability.

\sect{Mass operators in the non-Abelian case.} As we have seen,
there exist several ways of introducing nonlocal gauge invariant
mass operators in the Abelian case. In particular, the four mass
operators (\ref{op1}), (\ref{op2}), (\ref{op3}) and (\ref{op4})
turn out to be equivalent. Let us face now the more complex case
of non-Abelian gauge theories. Let us start by considering the
operator $A_{\min }^{2}$.

\subsection{Non-Abelian generalization of the operator $A_{\min }^{2}$.}
The operator $A_{\min }^{2}$ of expression (\ref{op2}) can be
generalized to the non-Abelian case by minimizing the functional
$\mathrm{Tr}\int d^{4}x\,A_{\mu }^{u}A_{\mu }^{u}$ along the gauge
orbit of $A_{\mu }$ \cite
{Semenov,Zwanziger:1990tn,Dell'Antonio:1989jn,Dell'Antonio:1991xt,vanBaal:1991zw, Gubarev:2000nz,Gubarev:2000eu}%
, namely
\begin{eqnarray}
A_{\min }^{2} &\equiv &\min_{\{u\}}\mathrm{Tr}\int d^{4}x\,A_{\mu
}^{u}A_{\mu }^{u}\;,
\nonumber \\
A_{\mu }^{u} &=&u^{\dagger }A_{\mu }u+\frac{i}{g}u^{\dagger }\partial _{\mu
}u\;.  \label{Amin0}
\end{eqnarray}
A few remarks are in order. Although the minimization procedure
along the gauge orbit of $A_{\mu }$ makes the operator $A_{\min
}^{2}$ gauge invariant, it should be underlined that the explicit
determination of the absolute minimum achieved by the functional
$\mathrm{Tr}\int d^{4}x\,A_{\mu }^{u}A_{\mu }^{u}$ is a highly
nontrivial task which, in practice, requires the resolution of the
issue of the Gribov copies. It has been proven that the operator
$\mathrm{Tr}\int d^{4}x\,A_{\mu }^{u}A_{\mu }^{u}$ achieves its
absolute minimum along the gauge orbit of $A_{\mu }$ \cite
{Semenov,Zwanziger:1990tn,Dell'Antonio:1989jn,Dell'Antonio:1991xt,vanBaal:1991zw}%
. Moreover, it is also known that, in general, it possesses many
relative minima along a given gauge orbit. Therefore, one has to
be sure that the correct minimum has been selected. This requires
a detailed knowledge of the so called fundamental modular region,
which is the set of all absolute minima in field space of the
functional $\mathrm{Tr}\int d^{4}x\,A_{\mu }^{u}A_{\mu }^{u}$. The
fundamental modular region is contained in the Gribov region,
which is defined as the set of all relative minima of
$\mathrm{Tr}\int d^{4}x\,A_{\mu }^{u}A_{\mu }^{u}$. While the
Gribov region turns out to be still plagued by the presence of
additional Gribov copies, the interior of the fundamental modular
region is free from Gribov copies \cite
{Semenov,Zwanziger:1990tn,Dell'Antonio:1989jn,Dell'Antonio:1991xt,vanBaal:1991zw, Gubarev:2000nz,Gubarev:2000eu}%
, a feature of primary importance for a correct quantization of
Yang-Mills theories. However, a knowledge of the fundamental
modular region of practical use in the Feynman path integral is
not yet at our disposal. All this should give to the reader an
idea of the real difficulty of obtaining an explicit expression
for the absolute minimum configuration of the functional
$\mathrm{Tr}\int d^{4}x\,A_{\mu }^{u}A_{\mu }^{u}$. A more modest
program would be that of considering the Gribov region instead of
the fundamental modular region, amounting to consider field
configurations which are relative minima of $\mathrm{Tr}\int
d^{4}x\,A_{\mu }^{u}A_{\mu }^{u}$. These configurations can be
constructed in a relatively easy way as formal power series in the
gauge field $A_{\mu }$. As discussed in Appendix \ref{apb}, a
minimum configuration of $\mathrm{Tr}\int d^{4}x\,A_{\mu
}^{u}A_{\mu }^{u}$ is attained when $u=h$ so that $A_{\mu }^{h}$
is a transverse field, $\partial _{\mu }A_{\mu }^{h}=0$. The
transversality condition can be solved order by order
\cite{Lavelle:1995ty}, allowing us to express $h$ as a formal
power series in the gauge field $A_{\mu }$, \textit{i.e.}
$h=h(A)$. This gives
\begin{eqnarray}
A_{\mu }^{h} &=&\left( \delta _{\mu \nu }-\frac{\partial _{\mu }\partial
_{\nu }}{\partial ^{2}}\right) \phi _{\nu }\;,  \nonumber \\
\phi _{\nu } &=&A_{\nu }-ig\left[ \frac{1}{\partial ^{2}}\partial A,A_{\nu
}\right] +\frac{ig}{2}\left[ \frac{1}{\partial ^{2}}\partial A,\partial
_{\nu }\frac{1}{\partial ^{2}}\partial A\right] +O(A^{3})\;.  \label{min0}
\end{eqnarray}
In particular, the configuration $A_{\mu }^{h}$ turns out to be invariant
under infinitesimal gauge transformations order by order in the gauge
coupling $g$ \cite{Lavelle:1995ty}, see also Appendix \ref{apb}, namely
\begin{eqnarray}
\delta A_{\mu }^{h} &=&0\;,  \nonumber \\
\delta A_{\mu } &=&-\partial _{\mu }\omega +ig\left[ A_{\mu },\omega \right]
\;.  \label{gio}
\end{eqnarray}
Thus, from expression (\ref{min0}) it follows that
\begin{eqnarray}
A_{\min }^{2} &=&\mathrm{Tr}\int d^{4}x\,A_{\mu }^{h}A_{\mu }^{h}\;,  \nonumber \\
&=&\frac{1}{2}\int d^{4}x\left[ A_{\mu }^{a}\left( \delta _{\mu \nu }-\frac{%
\partial _{\mu }\partial _{\nu }}{\partial ^{2}}\right) A_{\nu
}^{a}-gf^{abc}\left( \frac{\partial _{\nu }}{\partial ^{2}}\partial
A^{a}\right) \left( \frac{1}{\partial ^{2}}\partial {A}^{b}\right) A_{\nu
}^{c}\right] \;+O(A^{4})\;.  \label{min1}
\end{eqnarray}
We see that the operator $A_{\min }^{2}$ can be expressed as an infinite sum
of nonlocal terms. Such a nonlocal structure looks almost hopeless to be
handled in a consistent way for a generic choice of the gauge fixing term.
The only possibility here seems that of adopting the Landau gauge condition,
$\partial _{\mu }A_{\mu }^{a}=0$. In this case, all nonlocal terms in the
r.h.s. of equation (\ref{min1}) drop out, so that
\begin{equation}
A_{\min }^{2}=\frac{1}{2}\int d^{4}xA_{\mu }^{a}A_{\mu }^{a}\;\;\;\;\mathrm{%
in\;the\;Landau\;gauge}\;.  \label{Amin1}
\end{equation}
It is worth remarking that, as proven in \cite{Dudal:2002pq}, the
massive Yang-Mills action
\begin{equation}
S_{m}=\frac{1}{4}\int d^{4}xF_{\mu \nu }^{a}F_{\mu \nu }^{a}+\frac{m^{2}}{2}%
\int d^{4}xA_{\mu }^{a}A_{\mu }^{a}+\int d^{4}x\left( b^{a}\partial _{\mu
}A_{\mu }^{a}+\bar{c}^{a}\partial _{\mu }D_{\mu }^{ab}c^{b}\right) \;,
\label{mass0}
\end{equation}
where $b^{a}$ is the Lagrange multiplier enforcing the Landau condition, $%
\partial _{\mu }A_{\mu }^{a}=0$, and $\bar{c}^{a}$, $c^{a}$ are the
Faddeev-Popov ghosts, is multiplicatively renormalizable to all
orders of perturbation theory.

\noindent In summary, we have seen that the operator $A_{\min
}^{2}$ can be generalized to the non-Abelian case. In addition,
when treated as a formal power series in the gauge field $A_{\mu
}$, it has the pleasant property of reducing to the renormalizable
operator $\int d^{4}xA_{\mu }^{a}A_{\mu }^{a}$ in the Landau
gauge. \newline
\newline
We also recall that the operator $\int d^{4}xA_{\mu }^{a}A_{\mu
}^{a}$ turns out to be renormalizable to all orders of
perturbation theory in the more general class of the linear
covariant gauges \cite{Dudal:2003np}, a fact
which has made possible to give evidence of a nonvanishing condensate $%
\left\langle A_{\mu }^{a}A_{\mu }^{a}\right\rangle $ in these gauges \cite
{Dudal:2003by}. However, outside of the Landau gauge, the relationship
between $A_{\min }^{2}$ and $\int d^{4}xA_{\mu }^{a}A_{\mu }^{a}$ is lost,
so that a study of the nonlocal operator $A_{\min }^{2}$ becomes difficult.
The operator $A_{\min }^{2}$ lacks thus a simple computational framework
outside of the Landau gauge.

\subsection{Non-Abelian generalization of the operator $\int d^{4}xA_{\mu
}^{T}A_{\mu }^{T}$.}
The discussion of the previous section allows us to generalize the operator $%
\int d^{4}xA_{\mu }^{T}A_{\mu }^{T}$ in the non-Abelian case. In
fact, according to expression (\ref{min0}) \cite{Lavelle:1995ty},
see also Appendix \ref{apb}, it is possible to introduce a gauge
invariant non-Abelian transverse field. It follows thus that the
non-Abelian generalization of the mass operator $\int d^{4}xA_{\mu
}^{T}A_{\mu }^{T}$ is provided by expression (\ref{min1}). This
establishes the equivalence between the non-Abelian version of
$\int d^{4}xA_{\mu }^{T}A_{\mu }^{T}$ and the functional $A_{\min
}^{2}$ within the space of the formal power series. Moreover, the
operator $\int d^{4}xA_{\mu }^{T}A_{\mu }^{T}$ is plagued by the
same difficulties affecting $A_{\min }^{2}$.

\subsection{Non-Abelian generalization of the Stueckelberg term.}
The Stueckelberg term, eq.(\ref{op3}), can be promoted to the
non-Abelian case \cite{Ruegg:2003ps}, namely
\begin{equation}
\mathcal{O}_{S}=\mathrm{Tr}\int d^{4}x\left( A_{\mu
}-\frac{i}{g}U^{-1}\partial _{\mu }U\right) ^{2}\;,
\label{stueck0}
\end{equation}
with
\begin{equation}
U=e^{ig\phi ^{a}T^{a}}\;,  \label{U0}
\end{equation}
where $\{T^{a}\}$, $a=1,...,N^{2}-1$, denote the hermitian generators of the
gauge group $SU(N)$, and where $\phi ^{a}$ is a dimensionless scalar field
in the adjoint representation. As shown in Appendix \ref{apc}, expression (%
\ref{stueck0}) is left invariant by the gauge transformations
\begin{eqnarray}
A_{\mu } &\rightarrow &V^{-1}A_{\mu }V+\frac{i}{g}V^{-1}\partial _{\mu }V\;,
\nonumber \\
U &\rightarrow &UV\;.  \label{gauge4}
\end{eqnarray}
The resulting non-Abelian massive action
\begin{equation}
S_{S}=\frac{1}{2}\mathrm{Tr}\int d^{4}x\,F{_{\mu \nu }}F{_{\mu \nu }}+m^{2}\mathrm{Tr}\int {%
d^{4}x}\,\left( A_{\mu }-\frac{i}{g}U^{-1}\partial _{\mu }U\right) ^{2}\;,
\label{action2}
\end{equation}
looks local. However, it is not polynomial in the Stueckelberg field $\phi
^{a}$. In fact, when expanded in a power series in the field $\phi ^{a}$,
the term $U^{-1}\partial _{\mu }U$ gives rise to an infinite number of
vertices. This jeopardizes a consistent perturbative treatment of expression
(\ref{action2}). To the best of our knowledge, the action (\ref{action2}) is
not multiplicatively renormalizable \cite{Ruegg:2003ps}, see also the recent
discussion given in \cite{Ferrari:2004pd}. As done in the Abelian case, it
is interesting to have a look at the classical equations of motion which
follow from the action (\ref{action2}), \textit{i.e.}
\begin{equation}
D_{\mu }\left( A_{\mu }-\frac{i}{g}U^{-1}\partial _{\mu }U\right) =0\;.
\label{eq2}
\end{equation}
Equation (\ref{eq2}) can be used to express the Stueckelberg field $\phi ^{a}
$ as a power series in the gauge field $A_{\mu }^{a}$ \cite{Esole:2004rx}, see also Appendix \ref{apc}%
, yielding
\begin{equation}
\phi ^{a}=-\frac{1}{\partial ^{2}}\partial A^{a}+\frac{g}{\partial ^{2}}%
\left( f^{abc}A_{\mu }^{b}\partial _{\mu }\frac{\partial A^{c}}{\partial ^{2}%
}+\frac{1}{2}f^{abc}\partial A^{b}\frac{1}{\partial ^{2}}\partial
A^{c}\right) +O(A^{3})\;.  \label{stueck1}
\end{equation}
Therefore
\begin{eqnarray}
A_{\mu }-\frac{i}{g}U^{-1}\partial _{\mu }U &=&T^{a}\left[ A_{\mu
}^{a}-\partial _{\mu }\frac{\partial A^{a}}{\partial ^{2}}-\frac{g}{2}f^{abc}%
\frac{\partial _{\mu }}{\partial ^{2}}\partial A^{b}\frac{1}{\partial ^{2}}%
\partial A^{c}+\right.   \nonumber \\
&+&\left. \frac{g}{\partial ^{2}}\partial _{\mu }\left( f^{abc}A_{\nu
}^{b}\partial _{\nu }\frac{\partial A^{c}}{\partial ^{2}}+\frac{1}{2}%
f^{abc}\partial A^{b}\frac{1}{\partial ^{2}}\partial A{^{c}}\right) \right]
+O(A^{3})\;.  \label{ast}
\end{eqnarray}
Thus
\begin{eqnarray}
\mathcal{O}_{S} &=&\mathrm{Tr}\int d^{4}x\left( A_{\mu
}-\frac{i}{g}U^{-1}\partial
_{\mu }U\right) ^{2}  \nonumber \\
&=&\frac{1}{2}\int d^{4}x\left[ A_{\mu }^{aT}A_{\mu }^{aT}+2gA_{\mu }^{aT}%
\frac{\partial _{\mu }}{\partial ^{2}}\left( f^{abc}A_{\nu }^{b}\partial
_{\nu }\frac{\partial A^{c}}{\partial ^{2}}+\frac{1}{2}f^{abc}\partial A^{b}%
\frac{1}{\partial ^{2}}\partial A^{c}\right) \right.   \nonumber \\
&-&\left. gf^{abc}A_{\mu }^{aT}\left( \partial _{\mu }\frac{\partial A^{b}}{%
\partial ^{2}}\right) \frac{\partial A^{c}}{\partial ^{2}}\right]
+O(A^{4})\;.  \label{ast1}
\end{eqnarray}
Moreover, taking into account that, due to the transversality of $A_{\mu
}^{aT}$, the second term of the expression above vanishes by integration by
parts, we obtain
\begin{equation}
\mathcal{O}_{S}=\frac{1}{2}\mathrm{Tr}\int d^{4}x\left[ A_{\mu
}^{a}\left( \delta _{\mu \nu }-\frac{\partial _{\mu }\partial
_{\nu }}{\partial ^{2}}\right)
A_{\mu }^{a}-gf^{abc}A_{\mu }^{aT}\left( \partial _{\mu }\frac{\partial A^{b}%
}{\partial ^{2}}\right) \frac{\partial A^{c}}{\partial ^{2}}\right]
+O(A^{4})\;,  \label{ast2}
\end{equation}
which coincides precisely with expression (\ref{min1}). This shows
the classical equivalence, within the space of the formal power
series, between the Stueckelberg mass operator and the functional
$A_{\min }^{2}$ in the non-Abelian case.

\subsection{Non-Abelian generalization of the operator $\int d^{4}xF_{\mu \nu
}\frac{1}{\partial ^{2}}F_{\mu \nu }$.}
It remains now to discuss the non-Abelian generalization of the operator $%
\int d^{4}xF_{\mu \nu }\frac{1}{\partial ^{2}}F_{\mu \nu }$, a
task easily achieved by replacing the ordinary derivative,
$\partial $, by the covariant one, $D$, namely
\begin{equation}
 \mathrm{\mathrm{Tr}}\int d^{4}xF_{\mu \nu }\frac{1}{D^{2}}F_{\mu
\nu }\equiv \frac{1}{2} \int d^{4}xF_{\mu \nu
}^a\left[(D^{2})^{-1}\right]^{ab}F_{\mu \nu }^b\;. \label{op4a}
\end{equation}
We remark that this term can be introduced in any gauge and,
unlike the functional $A_{\min }^{2}$, does not require any
specific knowledge of the properties of the Gribov region as well
as of the fundamental modular region. It has already been
considered in \cite{Jackiw:1997jg} in the case of the
three-dimensional Yang-Mills theories, where the use of the
operator (\ref{op4a}) was based on its appearance in e.g. the
two-dimensional Schwinger model. However, so far, it has not yet
been analysed in four dimensions. Although in the Abelian case the
operator $\int d^{4}xF_{\mu \nu }\frac{1}{\partial ^{2}}F_{\mu \nu
}$ turns out to be equivalent to $A_{\min }^{2}$, this is no more
true in the non-Abelian case. This can be understood by observing
that, thanks to gauge invariance, the expression (\ref{min1}) for
$A_{\min }^{2}$ can be rewritten directly in terms of the field
strength $F_{\mu \nu }$. In fact, as proven in
\cite{Zwanziger:1990tn}, it turns out that
\begin{eqnarray}
A_{\min }^{2} &=&-\frac{1}{2}\mathrm{Tr}\int d^{4}x\left( F_{\mu
\nu }\frac{1}{D^{2}}F_{\mu \nu }+2i\frac{1}{D^{2}}F_{\lambda \mu
}\left[ \frac{1}{D^{2}}D_{\kappa }F_{\kappa \lambda
},\frac{1}{D^{2}}D_{\nu }F_{\nu \mu }\right] \right.
\nonumber \\
&&-2i\left. \frac{1}{D^{2}}F_{\lambda \mu }\left[ \frac{1}{D^{2}}D_{\kappa
}F_{\kappa \nu },\frac{1}{D^{2}}D_{\nu }F_{\lambda \mu }\right] \right)
+O(F^{4})\;,  \label{zzw}
\end{eqnarray}
from which the difference between the operator (\ref{op4a}) and $A_{\min
}^{2}$ becomes apparent. This interesting feature gives to the operator (\ref
{op4a}) a privileged role with respect to the localization procedure. In
fact, while in the case of $A_{\min }^{2}$ one has to deal with an infinite
number of nonlocal terms, expression (\ref{op4a}) seems to be more
manageable. In the next section the localization procedure of the operator (%
\ref{op4a}) will be discussed.

\sect{Localizing the mass operator $\mathrm{Tr}\int d^{4}xF_{\mu \nu }\frac{1}{%
D^{2}}F_{\mu \nu }$.}
The localization of the operator $\mathrm{Tr}$ $\int d^{4}xF_{\mu \nu }\frac{1}{D^{2}}%
F_{\mu \nu }$ can be achieved by generalizing the procedure adopted in the
localization of the Abelian operator $\int d^{4}xF_{\mu \nu }\frac{1}{%
\partial ^{2}}F_{\mu \nu }$, eq.$\left( \ref{g5}\right) $. Let us start by
considering the Yang-Mills action with the addition of the mass operator (%
\ref{op4a}), \textit{i.e. }
\begin{equation}
S_{YM}+S_{\mathcal{O}}\;,  \label{ymop}
\end{equation}
where
\begin{equation}
S_{YM}=\frac{1}{4}\int d^{4}xF_{\mu \nu }^{a}F_{\mu \nu }^{a}\;,  \label{ym}
\end{equation}
and
\begin{equation}
S_{\mathcal{O}}=-\frac{m^{2}}{4}\int d^{4}xF_{\mu \nu }^{a}\left[ \left(
D^{2}\right) ^{-1}\right] ^{ab}F_{\mu \nu }^{b}\;.  \label{massop}
\end{equation}
The term (\ref{massop}) can be localized by means of the introduction of a
pair of complex bosonic antisymmetric tensor fields in the adjoint
representation, $\left( B_{\mu \nu }^{a},\bar{B}_{\mu \nu }^{a}\right) $,
according to
\begin{equation}
e^{-S_{\mathcal{O}}}=\int D\bar{B}DB\left( \det D^{2}\right) ^{6}\exp \left[
-\left( \frac{1}{4}\int d^{4}x\bar{B}_{\mu \nu }^{a}D_{\sigma
}^{ab}D_{\sigma }^{bc}B_{\mu \nu }^{c}+\frac{im}{4}\int d^{4}x\left( B-\bar{B%
}\right) _{\mu \nu }^{a}F_{\mu \nu }^{a}\right) \right] \;,  \label{loc2}
\end{equation}
where the determinant, $\left( \det D^{2}\right) ^{6}$, takes into account
the Jacobian arising from the integration over the bosonic complex fields $%
\left( B_{\mu \nu }^{a},\bar{B}_{\mu \nu }^{a}\right) $. This term can also
be localized by means of suitable anticommuting antisymmetric tensor fields $%
\left( \bar{G}_{\mu \nu }^{a},G_{\mu \nu }^{a}\right) $, namely
\begin{equation}
\left( \det D^{2}\right) ^{6}=\int D\bar{G}DG\exp \left( \frac{1}{4}\int {%
d^{4}x}\bar{G}_{\mu \nu }^{a}D_{\sigma }^{ab}D_{\sigma }^{bc}G_{\mu \nu
}^{c}\right) \;.  \label{loc3}
\end{equation}
Therefore, we obtain a classical local action which reads
\begin{equation}
S_{YM}+S_{BG}+S_{m}\;,  \label{action1}
\end{equation}
where
\begin{eqnarray}
S_{BG} &=&\frac{1}{4}\int d^{4}x\left( \bar{B}_{\mu \nu }^{a}D_{\sigma
}^{ab}D_{\sigma }^{bc}B_{\mu \nu }^{c}-\bar{G}_{\mu \nu }^{a}D_{\sigma
}^{ab}D_{\sigma }^{bc}G_{\mu \nu }^{c}\right) \;,  \nonumber \\
S_{m} &=&\frac{im}{4}\int d^{4}x\left( B-\bar{B}\right) _{\mu \nu
}^{a}F_{\mu \nu }^{a}\;.  \label{actions2}
\end{eqnarray}
The localization procedure does not destroy the gauge invariance of the
resulting action. In fact, it is easily checked that expression (\ref
{action1}) is left invariant by the gauge transformations
\begin{eqnarray}
\delta A_{\mu }^{a} &=&-D_{\mu }^{ab}\omega ^{b}\;,  \nonumber \\
\delta B_{\mu \nu }^{a} &=&gf^{abc}\omega ^{b}B_{\mu \nu }^{c}\;,  \nonumber
\\
\delta \bar{B}_{\mu \nu }^{a} &=&gf^{abc}\omega ^{b}\bar{B}_{\mu \nu }^{c}\;,
\nonumber \\
\delta G_{\mu \nu }^{a} &=&gf^{abc}\omega ^{b}G_{\mu \nu }^{c}\;,  \nonumber
\\
\delta \bar{G}_{\mu \nu }^{a} &=&gf^{abc}\omega ^{b}\bar{G}_{\mu \nu }^{c}\;,
\label{gtm}
\end{eqnarray}
\begin{equation}
\delta \left( S_{YM}+S_{BG}+S_{m}\right) =0\;,  \label{gtminv}
\end{equation}
so that condition (I.) is fulfilled. Let us proceed thus with the
identification of a suitable quantized action, associated to
expression (\ref {action1}), which enjoys the property of being
multiplicatively renormalizable. For that, we follow the setup
successfully introduced by Zwanziger
\cite{Zwanziger:1989mf,Zwanziger:1992qr} in the localization of
the nonlocal horizon function implementing the restriction to the
Gribov region in the Landau gauge. In a series of papers,
Zwanziger has been able to show that the restriction to the Gribov
region can be implemented by adding to the Yang-Mills action a
nonlocal term, known as the horizon function, which is given by
\begin{equation}
S_{\mathrm{Horiz}}=\gamma ^{4}g^{2}\int d^{4}xf^{abc}A_{\mu }^{b}\left(
\mathcal{M}^{-1}\right) ^{ad}f^{dec}A_{\mu }^{e}\;,  \label{g9}
\end{equation}
where $\gamma $ denotes the Gribov parameter \cite{Gribov:1977wm} and $%
\mathcal{M}^{ab}$ is the Faddeev-Popov operator of the Landau gauge
\begin{equation}
\mathcal{M}^{ab}=-\partial _{\mu }\left( \partial _{\mu }\delta
^{ab}+gf^{acb}A_{\mu }^{c}\right) \;.  \label{g10}
\end{equation}
As proven in \cite{Zwanziger:1989mf,Zwanziger:1992qr}, the nonlocal horizon
term $\left( \ref{g9}\right) $ can be localized by means of a suitable set
of additional fields, in a way analogous to that of eq.$\left( \ref{loc2}%
\right) $. Remarkably, the resulting theory is renormalizable to all orders,
obeying the renormalization group equations. Thus, it seems natural to us to
adopt here the same procedure. According to \cite
{Zwanziger:1989mf,Zwanziger:1992qr}, we treat the operators $B_{\mu \nu
}^{a}F_{\mu \nu }^{a}\;$and $\bar{B}_{\mu \nu }^{a}F_{\mu \nu }^{a}$,
entering the expression for $S_{m}$ in eq.(\ref{actions2}), as composite
operators coupled to suitable external sources $V_{\sigma \rho \mu \nu }(x)$%
, $\bar{V}_{\sigma \rho \mu \nu }(x)$. This amounts to replace the term $%
S_{m}$ by
\begin{equation}
\frac{1}{4}\int d^{4}x\left( V_{\sigma \rho \mu \nu }\bar{B}_{\sigma \rho
}^{a}F_{\mu \nu }^{a}-\bar{V}_{\sigma \rho \mu \nu }B_{\sigma \rho
}^{a}F_{\mu \nu }^{a}\right) \;.  \label{rs}
\end{equation}
At the end, the sources $V_{\sigma \rho \mu \nu }(x)$, $\bar{V}_{\sigma \rho
\mu \nu }(x)$ are required to attain their physical value, namely
\begin{equation}
\bar{V}_{\sigma \rho \mu \nu }\Big|_{\mathrm{phys}}=V_{\sigma \rho \mu \nu }%
\Big|_{\mathrm{phys}}\;=\;\frac{-im}{2}\left( \delta _{\sigma \mu }\delta
_{\rho \nu }-\delta _{\sigma \nu }\delta _{\rho \mu }\right) \;,  \label{ps}
\end{equation}
so that expression (\ref{rs}) gives back the term $S_{m}$. As
pointed out in \cite{Zwanziger:1989mf,Zwanziger:1992qr}, this
procedure allows us to study the renormalization properties of the
Green's functions obtained from the action $\left(
S_{YM}+S_{BG}\right) $ with the insertion of the composite
operators $B_{\mu \nu }^{a}F_{\mu \nu }^{a}\;$and $\bar{B}_{\mu
\nu }^{a}F_{\mu \nu }^{a}$. Following
\cite{Zwanziger:1989mf,Zwanziger:1992qr}, let us focus first on
the properties of the action $\left( S_{YM}+S_{BG}\right) $ which,
as we shall see, displays a rich symmetry content.

\subsection{BRST invariance.}
In this section we shall discuss the symmetry content of the action
$\left(
S_{YM}+S_{BG}\right) $, where $S_{YM}$ is the Yang-Mills action, eq.(\ref{ym}%
), and $S_{BG}$ depends on the localizing fields $\left( B_{\mu \nu }^{a},%
\bar{B}_{\mu \nu }^{a},\bar{G}_{\mu \nu }^{a},G_{\mu \nu }^{a}\right) $, eq.(%
\ref{actions2}). Let us begin by introducing the gauge fixing term, chosen
here to be that of the linear covariant gauges, namely
\begin{equation}
S=S_{YM}+S_{BG}+S_{gf}\;,  \label{sbgf}
\end{equation}
with
\begin{equation}
S_{gf}=\int d^{4}x\left( \frac{\alpha }{2}b^{a}b^{a}+b^{a}\partial _{\mu
}A_{\mu }^{a}+\bar{c}^{a}\partial _{\mu }D_{\mu }^{ab}c^{b}\right) \;,
\label{lg}
\end{equation}
where $b^{a}$ is the Lagrange multiplier and $\bar{c}^{a},c^{a}$ stand for
the Faddeev-Popov ghosts. It turns out that the action $S$ is left invariant
by the following $BRST$\ transformation, \textit{i.e.}
\begin{eqnarray}
sA_{\mu }^{a} &=&-D_{\mu }^{ab}c^{b}\;,  \nonumber \\
sc^{a} &=&\frac{g}{2}f^{abc}c^{a}c^{b}\;,  \nonumber \\
sB_{\mu \nu }^{a} &=&gf^{abc}c^{b}B_{\mu \nu }^{c}+G_{\mu \nu }^{a}\;,
\nonumber \\
s\bar{B}_{\mu \nu }^{a} &=&gf^{abc}c^{b}\bar{B}_{\mu \nu }^{c}\;,
\nonumber \\
sG_{\mu \nu }^{a} &=&gf^{abc}c^{b}G_{\mu \nu }^{c}\;,  \nonumber \\
s\bar{G}_{\mu \nu }^{a} &=&gf^{abc}c^{b}\bar{G}_{\mu \nu }^{c}+\bar{B}_{\mu
\nu }^{a}\;,  \nonumber \\
s\bar{c}^{a} &=&b^{a} \;, \nonumber \\
sb^{a} &=&0 \;, \nonumber \\
s^{2} &=&0\;,  \label{bi}
\end{eqnarray}
and
\begin{equation}
sS=0\;.  \label{binv}
\end{equation}
This is easily verified by observing that the term $S_{BG}$ can be written
as a pure $BRST$\ variation, according to
\begin{equation}
S_{BG}=\frac{1}{4}s\int d^{4}x\bar{G}_{\mu \nu }^{a}D_{\sigma
}^{ab}D_{\sigma }^{bc}B_{\mu \nu }^{c}\;.  \label{spu}
\end{equation}
Of course, the same property holds for the gauge fixing term
\begin{equation}
S_{gf}=s\int d^{4}x\left( \frac{\alpha }{2}\bar{c}^{a}b^{a}+\bar{c}%
^{a}\partial _{\mu }A_{\mu }^{a}\right) \;.  \label{spgf}
\end{equation}
In addition to the $BRST\;$invariance, and in complete analogy with the
Zwanziger action \cite{Zwanziger:1989mf,Zwanziger:1992qr} implementing the
restriction to the Gribov horizon, the model displays a global invariance $%
U(f)$, $f=6$, expressed by
\begin{equation}
\mathcal{Q}_{\mu \nu \alpha \beta }S=0\;,  \label{zw1}
\end{equation}
where
\begin{equation}
\mathcal{Q}_{\mu \nu \alpha \beta }=\int d^{4}x\left( B_{\alpha \beta }^{a}%
\frac{\delta }{\delta B_{\mu \nu }^{a}}-\bar{B}_{\mu \nu }^{a}\frac{\delta }{%
\delta \bar{B}_{\alpha \beta }^{a}}+G_{\alpha \beta }^{a}\frac{\delta }{%
\delta G_{\mu \nu }^{a}}-\bar{G}_{\mu \nu }^{a}\frac{\delta }{\delta \bar{G}%
_{\alpha \beta }^{a}}\right) \;.  \label{zw2}
\end{equation}
The presence of the global invariance $U(f)$ , $f=6$, means that one can
make use of the composite index $i\equiv \{\mu \nu \}$, $i=(1,\ldots ,6)$.
Therefore, setting
\begin{equation}
\left( B_{i}^{a},\bar{B}_{i}^{a},G_{i}^{a},\bar{G}_{i}^{a}\right) =\frac{1}{2%
}\left( B_{\mu \nu }^{a},\bar{B}_{\mu \nu }^{a},G_{\mu \nu }^{a},\bar{G}%
_{\mu \nu }^{a}\right) \;,  \label{if}
\end{equation}
we get
\begin{equation}
S_{BG}=\int d^{4}x\left( \bar{B}_{i}^{a}D_{\mu }^{ab}D_{\mu }^{bc}B_{i}^{c}-%
\bar{G}_{i}^{a}D_{\mu }^{ab}D_{\mu }^{bc}G_{i}^{c}\right) \;,  \label{gb3}
\end{equation}
and for the symmetry generator
\begin{equation}
\mathcal{Q}_{ij}=\int d^{4}x\left( B_{i}^{a}\frac{\delta }{\delta B_{j}^{a}}-%
\bar{B}_{j}^{a}\frac{\delta }{\delta \bar{B}_{i}^{a}}+G_{i}^{a}\frac{\delta
}{\delta G_{j}^{a}}-\bar{G}_{j}^{a}\frac{\delta }{\delta \bar{G}_{i}^{a}}%
\right) \;.  \label{zw3}
\end{equation}
By means of the diagonal operator $\mathcal{Q}_{f}=\mathcal{Q}_{ii}$, the $i-
$valued fields turn out to possess an additional quantum number, displayed
in Table \ref{table1}, together with the dimension and the ghost number.
\begin{table}[t]
\centering
\begin{tabular}{|c|c|c|c|c|c|c|c|c|}
\hline
fields & $A$ & $c$ & $\bar{c}$ & $b$ & $B$ & $\bar{B}$ & $G$ & $\bar{G}$ \\
\hline
dimension & 1 & 0 & 2 & 2 & 1 & 1 & 1 & 1 \\
ghost number & 0 & 1 & $-1$ & 0 & 0 & 0 & 1 & $-1$ \\
$\mathcal{Q}_{f}$-charge & 0 & 0 & 0 & 0 & 1 & $-1$ & 1 & $-1$ \\ \hline
\end{tabular}
\caption{Dimension, ghost number and $\mathcal{Q}_{f}$-charge of the fields.}
\label{table1}
\end{table}
Besides the global $U(f)$, $f=6$, invariance, the action (\ref{sbgf})
possesses the following additional rigid symmetries
\begin{equation}
\mathcal{R}_{ij}^{(A)}S=0\;,  \label{rig0}
\end{equation}
where $A=\{1,2,3,4\}$ and
\begin{eqnarray}
\mathcal{R}_{ij}^{(1)} &=&\int d^{4}x\left( B_{i}^{a}\frac{\delta }{\delta
G_{j}^{a}}-\bar{G}_{j}^{a}\frac{\delta }{\delta \bar{B}_{i}^{a}}\right) \;,
\nonumber \\
\mathcal{R}_{ij}^{(2)} &=&\int d^{4}x\left( \bar{B}_{i}^{a}\frac{\delta }{%
\delta \bar{G}_{j}^{a}}+G_{j}^{a}\frac{\delta }{\delta B_{i}^{a}}\right) \;,
\nonumber \\
\end{eqnarray}
\begin{eqnarray}
\mathcal{R}_{ij}^{(3)} &=&\int d^{4}x\left( \bar{B}_{i}^{a}\frac{\delta }{%
\delta G_{j}^{a}}-\bar{G}_{j}^{a}\frac{\delta }{\delta B_{i}^{a}}\right) \;,
\nonumber \\
\mathcal{R}_{ij}^{(4)} &=&\int {d^{4}x}\left( B_{i}^{a}\frac{\delta }{\delta
\bar{G}_{j}^{a}}+G_{j}^{a}\frac{\delta }{\delta \bar{B}_{i}^{a}}\right) \;.
\label{oprig0}
\end{eqnarray}
Let us conclude this section by showing that also the source term
(\ref{rs}) can be introduced in a $BRST$ invariant way. This is
achieved by considering the following source term
\begin{eqnarray}
S_{aux} &=&s\int d^{4}x\left[ \left( V_{i\mu \nu }\bar{G}_{i}^{a}-\bar{U}%
_{i\mu \nu }B_{i}^{a}\right) F_{\mu \nu }^{a}+\chi _{1}\bar{U}_{i\mu \nu
}\partial ^{2}V_{i\mu \nu }\right.   \nonumber \\
 &+&\left.\chi _{2}\bar{U}_{i\mu \nu }\partial _{\mu }\partial _{\alpha
}V_{i\nu \alpha }-\zeta \left( \bar{U}_{i\mu \nu }V_{i\mu \nu }\bar{V}%
_{j\alpha \beta }V_{j\alpha \beta }-\bar{U}_{i\mu \nu }V_{i\mu \nu }\bar{U}%
_{j\alpha \beta }U_{j\alpha \beta }\right) \right] \;,  \label{ss}
\end{eqnarray}
with
\begin{eqnarray}
sV_{i\mu \nu } &=&U_{i\mu \nu }\;,  \nonumber \\
sU_{i\mu \nu } &=&0\;,  \nonumber \\
s\bar{U}_{i\mu \nu } &=&\bar{V}_{i\mu \nu }\;,  \nonumber \\
s\bar{V}_{i\mu \nu } &=&0\;,  \nonumber \\
s^{2} &=&0\;.  \label{sbrst}
\end{eqnarray}
The quantum numbers of the sources are displayed in Table
\ref{table2}. Therefore, for $S_{aux}$ one gets
\begin{eqnarray}
S_{aux} &=&\int d^{4}x\left[ \bar{U}_{i\mu \nu }G_{i}^{a}F_{\mu \nu
}^{a}+V_{i\mu \nu }\bar{B}_{i}^{a}F_{\mu \nu }^{a}-\bar{V}_{i\mu \nu
}B_{i}^{a}F_{\mu \nu }^{a}+U_{i\mu \nu }\bar{G}_{i}^{a}F_{\mu \nu }^{a}+\chi
_{1}\left( \bar{V}_{i\mu \nu }\partial ^{2}V_{i\mu \nu }\right. \right.
\nonumber \\
&-&\left. \left. \bar{U}_{i\mu \nu }\partial ^{2}U_{i\mu \nu
}\right) +\chi _{2}\left( \bar{V}_{i\mu \nu }\partial _{\mu
}\partial _{\alpha }V_{i\nu \alpha }-\bar{U}_{i\mu \nu }\partial
_{\mu }\partial _{\alpha }{U}_{i\nu
\alpha }\right) -\zeta \left( \bar{U}_{i\mu \nu }U_{i\mu \nu }\bar{U}%
_{j\alpha \beta }U_{j\alpha \beta }\right. \right.   \nonumber \\
&+&\left. \left. \bar{V}_{i\mu \nu }V_{i\mu \nu }\bar{V}_{j\alpha
\beta }V_{j\alpha \beta }-2\bar{U}_{i\mu \nu }U_{i\mu \nu
}\bar{V}_{j\alpha \beta }V_{j\alpha \beta }\right) \right] \;.
\label{ssex}
\end{eqnarray}
The parameters $\chi _{1}$, $\chi _{2}$ and $\zeta $ are free parameters,
needed for renormalizability purposes. The action $S_{aux}$ reduces to the
term $S_{m}$ of eq.(\ref{actions2}) when the sources $\left( V_{i\mu \nu },%
\bar{V}_{i\mu \nu },U_{i\mu \nu },\bar{U}_{i\mu \nu }\right) $ attain their
physical values, given now by
\begin{equation}
\left( V_{i\mu \nu },\bar{V}_{i\mu \nu },U_{i\mu \nu },\bar{U}_{i\mu \nu
}\right) =\frac{1}{2}\left( V_{\sigma \rho \mu \nu },\bar{V}_{\sigma \rho
\mu \nu },U_{\sigma \rho \mu \nu },\bar{U}_{\sigma \rho \mu \nu }\right) \;,
\label{1q}
\end{equation}
\begin{eqnarray}
\bar{V}_{\sigma \rho \mu \nu }\Big|_{\mathrm{phys}} &=&V_{\sigma \rho \mu
\nu }\Big|_{\mathrm{phys}}\;=\;\frac{-im}{2}\left( \delta _{\sigma \mu
}\delta _{\rho \nu }-\delta _{\sigma \nu }\delta _{\rho \mu }\right) \;,
\nonumber \\
U_{\sigma \rho \mu \nu } &=&\bar{U}_{\sigma \rho \mu \nu }=0\;.  \label{2q}
\end{eqnarray}
Thus
\begin{equation}
S_{aux}\bigg|_{\mathrm{phys}}\rightarrow S_{m}-\frac{9}{4}\int d^{4}x{%
\;\zeta }m{^{4}}\;,\;  \label{3q}
\end{equation}
so that the term $S_{m}$ is recovered, modulo the constant quantity ${\zeta }%
m{^{4}}$. All ingredients needed to study the renormalizability of the
action
\begin{equation}
S_{YM}+S_{BG}+S_{gf}+S_{aux}\;,  \label{ca}
\end{equation}
are now at our disposal. This will be the task of the next section.
\begin{table}[t]
\centering
\begin{tabular}{|c|c|c|c|c|}
\hline sources & $\bar{U}$ & $V$ & $U$ & $\bar{V}$ \\ \hline
dimension & 1 & 1 & 1 & 1 \\
ghost number & $-1$ & 0 & 1 & 0 \\
$\mathcal{Q}_{f}$-charge & $-1$ & 1 & 1 & $-1$ \\ \hline
\end{tabular}
\caption{Dimension, ghost number and $\mathcal{Q}_{f}$-charge of
the sources. } \label{table2}
\end{table}

\sect{Identification of a multiplicatively renormalizable action.}
In order to discuss the renormalizability properties of our model,
we have first to write down all possible Ward identities expressing
the symmetry content of the starting classical action,
eq.(\ref{ca}). Let us begin by working out the Slavnov-Taylor
identity. Following the algebraic renormalization procedure as
described in \cite{Piguet:1995er}, we need to introduce
additional external sources $\left( \Omega _{\mu }^{a},L^{a},\bar{Y}%
_{i}^{a},Y_{i}^{a},\bar{X}_{i}^{a},X_{i}^{a}\right) $ in order to define at
the quantum level the composite operators entering the nonlinear $BRST$
transformations of the fields $\left( A_{\mu }^{a},c^{a},B_{i}^{a},\bar{B}%
_{i}^{a},G_{i}^{a},\bar{G}_{i}^{a}\right) $, eqs.(\ref{bi}). In
the present case, this term reads
\begin{equation}
S_{ext}=s\int {d^{4}x}\left( -\Omega _{\mu }^{a}A_{\mu }^{a}+L^{a}c^{a}-\bar{%
Y}_{i}^{a}B_{i}^{a}-Y_{i}^{a}\bar{B}_{i}^{a}+\bar{X}%
_{i}^{a}G_{i}^{a}+X_{i}^{a}\bar{G}_{i}^{a}\right) \;,  \label{ext}
\end{equation}
with
\begin{equation}
s\Omega _{\mu }^{a}=sL^{a}=0\;,  \label{e1}
\end{equation}
and
\begin{eqnarray}
sY_{i}^{a} &=&X_{i}^{a}  \nonumber \\
sX_{i}^{a} &=&0 \\
s\bar{X}_{i}^{a} &=&-\bar{Y}_{i}^{a} \\
s\bar{Y}_{i}^{a} &=&0\;.  \label{e2}
\end{eqnarray}
The quantum numbers of the external sources $\left( \Omega _{\mu }^{a},L^{a},%
\bar{Y}_{i}^{a},Y_{i}^{a},\bar{X}_{i}^{a},X_{i}^{a}\right) $ are displayed
in Table \ref{table3}.
\begin{table}[t]
\centering
\begin{tabular}{|c|c|c|c|c|c|c|}
\hline
sources & $\Omega$ & $L$ & $\bar{Y}$ & $Y$ & $\bar{X}$ & $X$ \\ \hline
dimension & 3 & 4 & 3 & 3 & 3 & 3 \\
ghost number & $-1$ & $-2$ & $-1$ & $-1$ & $-2$ & 0 \\
$\mathcal{Q}_{f}$-charge & 0 & 0 & $-1$ & 1 & $-1$ & 1 \\ \hline
\end{tabular}
\caption{Dimension, fermion number and $\mathcal{Q}_{f}$-charge of the
external sources.}
\label{table3}
\end{table}
For the complete action $\Sigma $%
\begin{equation}
\Sigma =S_{YM}+S_{gf}+S_{BG}+S_{aux}+S_{ext}\;,  \label{as}
\end{equation}
we obtain
\begin{eqnarray}
\Sigma  &=&S_{YM}+\int d^{4}x\;\left( \frac{\alpha }{2}b^{a}b^{a}+b^{a}%
\partial _{\mu }A_{\mu }^{a}+\bar{c}^{a}\partial _{\mu }D_{\mu
}^{ab}c^{b}\right) +\int {d^{4}x}\left( \bar{B}_{i}^{a}D_{\mu }^{ab}D_{\mu
}^{bc}B_{i}^{c}-\bar{G}_{i}^{a}D_{\mu }^{ab}D_{\mu }^{bc}G_{i}^{c}\right)
\nonumber \\
&+&\int d^{4}x\left( \left( \bar{U}_{i\mu \nu }G_{i}^{a}+V_{i\mu \nu }\bar{B}%
_{i}^{a}-\bar{V}_{i\mu \nu }B_{i}^{a}+U_{i\mu \nu }\bar{G}_{i}^{a}\right)
F_{\mu \nu }^{a}+\chi _{1}\left( \bar{V}_{i\mu \nu }\partial ^{2}V_{i\mu \nu
}-\bar{U}_{i\mu \nu }\partial ^{2}U_{i\mu \nu }\right) \right)   \nonumber \\
&+&\int d^{4}x\chi _{2}\left( \bar{V}_{i\mu \nu }\partial _{\mu
}\partial _{\alpha }V_{i\nu \alpha }-\bar{U}_{i\mu \nu }\partial
_{\mu }\partial _{\alpha }U_{i\nu \alpha }\right) -\int
d^{4}x\zeta \left( \bar{U}_{i\mu \nu }U_{i\mu \nu
}\bar{U}_{j\alpha \beta }U_{j\alpha \beta }+\bar{V}_{i\mu \nu
}V_{i\mu \nu }\bar{V}_{j\alpha \beta }V_{j\alpha \beta
}\right.\nonumber\\&-&\left.2\bar{U}_{i\mu \nu }U_{i\mu \nu
}\bar{V}_{j\alpha \beta }V_{j\alpha \beta }\right)  +\int d^{4}x\left( -\Omega _{\mu }^{a}D_{\mu }^{ab}c^{b}+\frac{g}{2}%
f^{abc}L^{a}c^{b}c^{c}+gf^{abc}\bar{Y}%
_{i}^{a}c^{b}B_{i}^{c}+gf^{abc}Y_{i}^{a}c^{b}\bar{B}_{i}^{c}\right.
\nonumber \\
 &+&\left.gf^{abc}\bar{X}_{i}^{a}c^{b}G_{i}^{c}+gf^{abc}X_{i}^{a}c^{b}\bar{G}%
_{i}^{c}\right) \;.  \label{ass1}
\end{eqnarray}
Expression (\ref{ass1}) obeys several Ward identities, which we
enlist below
\begin{itemize}
\item  the Slavnov-Taylor identity
\begin{equation}
\mathcal{S}(\Sigma )=0\;,  \label{sti}
\end{equation}
\end{itemize}
\begin{eqnarray}
\mathcal{S}(\Sigma ) &=&\int d^{4}x\left[ \frac{\delta \Sigma }{\delta
\Omega _{\mu }^{a}}\frac{\delta \Sigma }{\delta A_{\mu }^{a}}+\frac{\delta
\Sigma }{\delta L^{a}}\frac{\delta \Sigma }{\delta c^{a}}+b^{a}\frac{\delta
\Sigma }{\delta \bar{c}^{a}}+\left( \frac{\delta \Sigma }{\delta \bar{Y}%
_{i}^{a}}+G_{i}^{a}\right) \frac{\delta \Sigma }{\delta B_{i}^{a}}+\frac{%
\delta \Sigma }{\delta Y_{i}^{a}}\frac{\delta \Sigma }{\delta \bar{B}_{i}^{a}%
}+\frac{\delta \Sigma }{\delta \bar{X}_{i}^{a}}\frac{\delta \Sigma }{\delta {%
G}_{i}^{a}}\right.  \nonumber \\
&+&\left. \left( \frac{\delta \Sigma }{\delta X_{i}^{a}}+\bar{B}%
_{i}^{a}\right) \frac{\delta \Sigma }{\delta \bar{G}_{i}^{a}}+\bar{V}_{i\mu
\nu }\frac{\delta \Sigma }{\delta \bar{U}_{i\mu \nu }}+U_{i\mu \nu }\frac{%
\delta \Sigma }{\delta {V}_{i\mu \nu }}-\bar{Y}_{i}^{a}\frac{\delta \Sigma }{%
\delta \bar{X}_{i}^{a}}+X_{i}^{a}\frac{\delta \Sigma }{\delta Y_{i}^{a}}%
\right] \;,  \label{stie}
\end{eqnarray}
\begin{itemize}
\item  the global $U(f)$ invariance , $f=6$, \textit{i.e.}
\begin{equation}
\mathcal{Q}_{ij}\Sigma =0\;,  \label{zw5}
\end{equation}
where
\begin{eqnarray}
\mathcal{Q}_{ij} &=&\int d^{4}x\left( B_{i}^{a}\frac{\delta }{\delta
B_{j}^{a}}-\bar{B}_{j}^{a}\frac{\delta }{\delta \bar{B}_{i}^{a}}+G_{i}^{a}%
\frac{\delta }{\delta G_{j}^{a}}-\bar{G}_{j}^{a}\frac{\delta }{\delta \bar{G}%
_{i}^{a}}+U_{i\mu \nu }\frac{\delta }{\delta U_{j\mu \nu }}-\bar{U}_{j\mu
\nu }\frac{\delta }{\delta \bar{U}_{i\mu \nu }}\right.   \nonumber \\
&+&\left. V_{i\mu \nu }\frac{\delta }{\delta {V}_{j\mu \nu }}-\bar{V}_{j\mu
\nu }\frac{\delta }{\delta \bar{V}_{i\mu \nu }}+Y_{i}^{a}\frac{\delta }{%
\delta Y_{j}^{a}}-\bar{Y}_{j}^{a}\frac{\delta }{\delta \bar{Y}_{i}^{a}}%
+X_{i}^{a}\frac{\delta }{\delta X_{j}^{a}}-\bar{X}_{j}^{a}\frac{\delta }{%
\delta \bar{X}_{i}^{a}}\right) \;,  \label{zw6}
\end{eqnarray}

\item  the exact rigid symmetries
\begin{equation}
\mathcal{R}_{ij}^{(A)}\Sigma =0\;,  \label{rig2}
\end{equation}
where $A=\{1,2,3,4\}$ and
\begin{eqnarray}
\mathcal{R}_{ij}^{(1)} &=&\int d^{4}x\left( B_{i}^{a}\frac{\delta }{\delta
G_{j}^{a}}-\bar{G}_{j}^{a}\frac{\delta }{\delta \bar{B}_{i}^{a}}+V_{i\mu \nu
}\frac{\delta }{\delta U_{j\mu \nu }}-\bar{U}_{j\mu \nu }\frac{\delta }{%
\delta \bar{V}_{i\mu \nu }}+Y_{i}^{a}\frac{\delta }{\delta X_{j}^{a}}+\bar{X}%
_{j}^{a}\frac{\delta }{\delta \bar{Y}_{i}^{a}}\right) \;,  \nonumber \\
\mathcal{R}_{ij}^{(2)} &=&\int d^{4}x\left( \bar{B}_{i}^{a}\frac{\delta }{%
\delta \bar{G}_{j}^{a}}+G_{j}^{a}\frac{\delta }{\delta B_{i}^{a}}+\bar{V}%
_{i\mu \nu }\frac{\delta }{\delta \bar{U}_{j\mu \nu }}+U_{j\mu \nu }\frac{%
\delta }{\delta V_{i\mu \nu }}-\bar{Y}_{i}^{a}\frac{\delta }{\delta \bar{X}%
_{j}^{a}}+X_{j}^{a}\frac{\delta }{\delta Y_{i}^{a}}\right) \;,  \nonumber \\
\mathcal{R}_{ij}^{(3)} &=&\int d^{4}x\left( \bar{B}_{i}^{a}\frac{\delta }{%
\delta G_{j}^{a}}-\bar{G}_{j}^{a}\frac{\delta }{\delta B_{i}^{a}}-\bar{V}%
_{i\mu \nu }\frac{\delta }{\delta U_{j\mu \nu }}+\bar{U}_{j\mu \nu }\frac{%
\delta }{\delta V_{i\mu \nu }}+\bar{Y}_{i}^{a}\frac{\delta }{\delta X_{j}^{a}%
}+\bar{X}_{j}^{a}\frac{\delta }{\delta Y_{i}^{a}}\right) \;,  \nonumber \\
\mathcal{R}_{ij}^{(4)} &=&\int d^{4}x\left( B_{i}^{a}\frac{\delta }{\delta
\bar{G}_{j}^{a}}+G_{j}^{a}\frac{\delta }{\delta \bar{B}_{i}^{a}}-V_{i\mu \nu
}\frac{\delta }{\delta \bar{U}_{j\mu \nu }}-U_{j\mu \nu }\frac{\delta }{%
\delta \bar{V}_{i\mu \nu }}-Y_{i}^{a}\frac{\delta }{\delta \bar{X}_{j}^{a}}%
+X_{j}^{a}\frac{\delta }{\delta \bar{Y}_{i}^{a}}\right) \;,  \nonumber \\
&&  \label{rig3}
\end{eqnarray}

\item  the gauge fixing condition
\begin{equation}
\frac{\delta \Sigma }{\delta b^{a}}=\alpha b^{a}+\partial _{\mu }A{_{\mu
}^{a}}\;  \label{b}
\end{equation}

\item  the antighost Ward identity
\begin{equation}
\bar{\mathcal{G}}^{a}\Sigma =0\;,  \label{antgh}
\end{equation}
where
\begin{equation}
\bar{\mathcal{G}}^{a}=\frac{\delta }{\delta \bar{c}^{a}}+\partial _{\mu }%
\frac{\delta }{\delta \Omega _{\mu }^{a}}\;.  \label{agh}
\end{equation}
\end{itemize}

\subsection{Determination of the most general local invariant counterterm}
Having established all the Ward identities fulfilled by the complete action $%
\Sigma $, we can now turn to the characterization of the most general
allowed counterterm $\Sigma ^{c}$. Following the algebraic renormalization
procedure \cite{Piguet:1995er}, $\Sigma ^{c}$ is an integrated local
polynomial in the fields and sources with dimension bounded by four, with
vanishing ghost number and $\mathcal{Q}_{f}$-charge, obeying the following
constraints
\begin{eqnarray}
\mathcal{Q}_{ij}\Sigma ^{c} &=&0\;,  \nonumber \\
\mathcal{R}_{ij}^{(A)}\Sigma ^{c} &=&0\;,  \nonumber \\
\frac{\delta \Sigma ^{c}}{\delta b^{a}} &=&0\;,  \nonumber \\
\bar{\mathcal{G}}^{a}\Sigma ^{c} &=&0\;,  \label{cc1}
\end{eqnarray}
in addition to
\begin{equation}
\mathcal{B}_{\Sigma }\Sigma ^{c}=0\;,  \label{cc2}
\end{equation}
where $\mathcal{B}_{\Sigma }$ is the nilpotent linearized Slavnov-Taylor
operator
\begin{eqnarray}
\mathcal{B}_{\Sigma } &=&\int d^{4}x\left[ \frac{\delta \Sigma }{\delta
\Omega _{\mu }^{a}}\frac{\delta }{\delta A_{\mu }^{a}}+\frac{\delta \Sigma }{%
\delta A_{\mu }^{a}}\frac{\delta }{\delta \Omega _{\mu }^{a}}+\frac{\delta
\Sigma }{\delta L^{a}}\frac{\delta }{\delta {c}^{a}}+\frac{\delta \Sigma }{%
\delta {c}^{a}}\frac{\delta }{\delta L^{a}}+b^{a}\frac{\delta }{\delta \bar{c%
}^{a}}+\left( \frac{\delta \Sigma }{\delta \bar{Y}_{i}^{a}}+G_{i}^{a}\right)
\frac{\delta }{\delta B_{i}^{a}}\right.   \nonumber \\
&+&\left. \frac{\delta \Sigma }{\delta B_{i}^{a}}\frac{\delta }{\delta \bar{Y%
}_{i}^{a}}+\frac{\delta \Sigma }{\delta Y_{i}^{a}}\frac{\delta }{\delta \bar{%
B}_{i}^{a}}+\left( \frac{\delta \Sigma }{\delta \bar{B}_{i}^{a}}%
+X_{i}^{a}\right) \frac{\delta }{\delta Y_{i}^{a}}+\frac{\delta \Sigma }{%
\delta \bar{X}_{i}^{a}}\frac{\delta }{\delta G_{i}^{a}}+\left( \frac{\delta
\Sigma }{\delta G_{i}^{a}}-\bar{Y}_{i}^{a}\right) \frac{\delta }{\delta \bar{%
X}_{i}^{a}}\right.   \nonumber \\
&+&\left. \left( \frac{\delta \Sigma }{\delta X_{i}^{a}}+\bar{B}%
_{i}^{a}\right) \frac{\delta }{\delta \bar{G}_{i}^{a}}+\frac{\delta \Sigma }{%
\delta \bar{G}_{i}^{a}}\frac{\delta }{\delta X_{i}^{a}}+\bar{V}_{i\mu \nu }%
\frac{\delta }{\delta \bar{U}_{i\mu \nu }}+U_{i\mu \nu }\frac{\delta }{%
\delta V_{i\mu \nu }}\right] \;,  \label{ncc2}
\end{eqnarray}
\begin{equation}
\mathcal{B}_{\Sigma }\mathcal{B}_{\Sigma }=0\;.  \label{nb}
\end{equation}
After a rather lengthy analysis, for the most general allowed counterterm we
have found
\begin{eqnarray}
\Sigma ^{c} &=&a_{0}S_{YM} + a_{1} \int d^{4}x
A_{\mu}^{a}\frac{\delta S_{YM}}{\delta {A}_{\mu
}^{a}}  \nonumber \\
&+&\int {d^{4}x}\left( \left( a_{1}+a_{2}\right) \left( \Omega _{\mu }^{a}+\partial _{\mu }%
\bar{c}^{a}\right) \partial _{\mu }{c}^{a}+a_{2}gf^{abc}\left( \Omega _{\mu
}^{a}+\partial _{\mu }\bar{c}^{a}\right) A_{\mu }^{b}c^{c}-a_{2}\frac{g}{2}%
f^{abc}L^{a}c^{b}c^{c}\right)   \nonumber \\
&+&\int {d^{4}x}\left\{ \left( 2a_{3}+a_{4}\right)
\bar{B}_{i}^{a}\partial ^{2}B_{i}^{a}-\left( 2a_{3}+a_{4}\right)
\bar{G}_{i}^{a}\partial
^{2}G_{i}^{a}\right.   \nonumber \\
&-&\left. \left( a_{1}+2a_{3}+a_{4}\right)
gf^{abc}\bar{B}_{i}^{a}\left(
\partial _{\mu }{A}_{\mu }^{b}+2A_{\mu }^{b}\partial _{\mu }\right)
B_{i}^{c}+\left( 2a_{1}+2a_{3}+a_{4}\right) g^{2}f^{abd}f^{bce}\bar{B}%
_{i}^{a}A_{\mu }^{d}A_{\mu }^{e}B_{i}^{c}\right.   \nonumber \\
&+&\left. \left( a_{1}+2a_{3}+a_{4}\right)
gf^{abc}\bar{G}_{i}^{a}\left(
\partial _{\mu }{A}_{\mu }^{b}+2A_{\mu }^{b}\partial _{\mu }\right)
G_{i}^{c}-\left( 2a_{1}+2a_{3}+a_{4}\right) g^{2}f^{abd}f^{bce}\bar{G}%
_{i}^{a}A_{\mu }^{d}A_{\mu }^{e}G_{i}^{c}\right.   \nonumber \\
&-&\left. a_{2}gf^{abc}c^{a}\left( \bar{Y}_{i}^{b}B_{i}^{c}+Y_{i}^{b}\bar{B}%
_{i}^{c}-\bar{X}_{i}^{b}G_{i}^{c}-X_{i}^{b}\bar{G}_{i}^{c}\right) +\left[
\left( a_{1}+a_{3}+a_{5}\right) 2\partial _{\mu }{A}_{\nu }^{a}\right.
\right.   \nonumber \\
&+&\left. \left. \left( 2a_{1}+a_{3}+a_{5}\right) gf^{abc}A_{\mu
}^{b}A_{\nu
}^{c}\right] \left( \bar{U}_{i\mu \nu }G_{i}^{a}+V_{i\mu \nu }\bar{B}%
_{i}^{a}+U_{i\mu \nu }\bar{G}_{i}^{a}-\bar{V}_{i\mu \nu }B_{i}^{a}\right)
\right.   \nonumber \\
&+&\left. \frac{\lambda^{abcd}}{16}\left( \bar{B}_{i}^{a}B_{i}^{b}-\bar{G}_{i}^{a}G_{i}^{b}%
\right)\left( \bar{B}_{j}^{c}B_{j}^{d}-\bar{G}_{j}^{c}G_{j}^{d}%
\right)+a_{7}\left( \bar{B}_{i}^{a}B_{i}^{a}-\bar{G}_{i}^{a}G_{i}^{a}%
\right) \left( \bar{V}_{i\mu \nu }V_{i\mu \nu }-\bar{U}_{i\mu \nu
}U_{i\mu
\nu }\right) \right.   \nonumber \\
&+&\left. a_{8}\left( \bar{B}_{i}^{a}G_{j}^{a}V_{i\mu \nu
}\bar{U}_{j\mu \nu
}+\bar{G}_{i}^{a}G_{j}^{a}U_{i\mu \nu }\bar{U}_{j\mu \nu }+\bar{B}%
_{i}^{a}B_{j}^{a}V_{i\mu \nu }\bar{V}_{j\mu \nu }-\bar{G}%
_{i}^{a}B_{j}^{a}U_{i\mu \nu }\bar{V}_{j\mu \nu }\right. \right.   \nonumber
\\
&-&\left. \left. G_{i}^{a}B_{j}^{a}\bar{U}_{i\mu \nu }\bar{V}_{j\mu \nu }+%
\bar{G}_{i}^{a}\bar{B}_{j}^{a}U_{i\mu \nu }V_{j\mu \nu }-\frac{1}{2}%
B_{i}^{a}B_{j}^{a}\bar{V}_{i\mu \nu }\bar{V}_{j\mu \nu }+\frac{1}{2}%
G_{i}^{a}G_{j}^{a}\bar{U}_{i\mu \nu }\bar{U}_{j\mu \nu }\right. \right.
\nonumber \\
&-&\left. \left. \frac{1}{2}\bar{B}_{i}^{a}\bar{B}_{j}^{a}V_{i\mu
\nu }V_{j\mu \nu }+\frac{1}{2}\bar{G}_{i}^{a}\bar{G}_{j}^{a}U_{i\mu
\nu }U_{j\mu
\nu }\right) +a_{9}\zeta \left( \bar{V}_{i\mu \nu }V_{i\mu \nu }-\bar{U}%
_{i\mu \nu }U_{i\mu \nu }\right) ^{2}\right.   \nonumber \\
&+&\left. a_{10}\chi _{1}\left( \bar{V}_{i\mu \nu }\partial
^{2}V_{i\mu \nu }-\bar{U}_{i\mu \nu }\partial ^{2}U_{i\mu \nu
}\right) +a_{11}\chi _{1}\left( \bar{V}_{i\mu \nu }\partial _{\mu
}\partial _{\alpha }{V}_{i\nu \alpha }-\bar{U}_{i\mu \nu }\partial
_{\mu }\partial _{\alpha }{U}_{i\nu \alpha }\right) \right\} \;,
\label{count2}
\end{eqnarray}
where $\left(
a_{0},a_{1},a_{2},a_{3},a_{4},a_{5},a_{7},a_{8},a_{9},a_{10},a_{11}
\right)$ are free parameters and $\lambda^{abcd}$ is an invariant
tensor of rank four with indices in the adjoint representation and
such that
\begin{eqnarray}
\lambda^{abcd}=\lambda^{cdab} \;, \nonumber \\
\lambda^{abcd}=\lambda^{bacd} \;. \label{abcd}
\end{eqnarray}
For a general discussion of the properties of higher rank
invariant tensors we refer the reader to
\cite{vanRitbergen:1998pn}. Let us only mention that an invariant
rank 4 tensor like $\lambda^{abcd}$ obeys a generalized Jacobi
identity
\begin{equation}\label{jacobigen}
    f^{man}\lambda^{mbcd}+f^{mbn}\lambda^{amcd}+f^{mcn}\lambda^{abmd}+f^{mdn}\lambda^{abcm}=0\,.
\end{equation}
These parameters $a_{i}$, $i=0,\ldots,11$, should correspond to a
multiplicative renormalization of the fields, parameters and
sources of the starting classical action $\Sigma $.\ However, it
turns out that the counterterm (\ref{count2}) cannot be reabsorbed
through a renormalization of
the parameters and fields of $\Sigma $. This means that the starting action $%
\Sigma $ is not stable against radiative corrections. Said otherwise, $%
\Sigma $ is not the most general local invariant action compatible
with the Ward identities (\ref{sti})-(\ref{antgh}). In fact, from
the expression (\ref{count2}) it follows that the term
\begin{eqnarray}
&& \int d^{4}x \left[a_{7}\left( \bar{B}_{i}^{a}B_{i}^{a}-\bar{G}%
_{i}^{a}G_{i}^{a}\right) \left( \bar{V}_{j\mu \nu }V_{j\mu \nu }-\bar{U}%
_{j\mu \nu }U_{j\mu \nu }\right) + \frac{\lambda^{abcd}}{16}\left( \bar{B}_{i}^{a}B_{i}^{b}-\bar{G}_{i}^{a}G_{i}^{b}%
\right)\left( \bar{B}_{j}^{c}B_{j}^{d}-\bar{G}_{j}^{c}G_{j}^{d}%
\right) \right.   \nonumber \\
&+&\left. a_{8}\left( \bar{B}_{i}^{a}G_{j}^{a}V_{i\mu \nu
}\bar{U}_{j\mu \nu
}+\bar{G}_{i}^{a}G_{j}^{a}U_{i\mu \nu }\bar{U}_{j\mu \nu }+\bar{B}%
_{i}^{a}B_{j}^{a}V_{i\mu \nu }\bar{V}_{j\mu \nu }-\bar{G}%
_{i}^{a}B_{j}^{a}U_{i\mu \nu }\bar{V}_{j\mu \nu }\right. \right.   \nonumber
\\
&-&\left. \left. G_{i}^{a}B_{j}^{a}\bar{U}_{i\mu \nu }\bar{V}_{j\mu \nu }+%
\bar{G}_{i}^{a}\bar{B}_{j}^{a}U_{i\mu \nu }V_{j\mu \nu }-\frac{1}{2}%
B_{i}^{a}B_{j}^{a}\bar{V}_{i\mu \nu }\bar{V}_{j\mu \nu }+\frac{1}{2}%
G_{i}^{a}G_{j}^{a}\bar{U}_{i\mu \nu }\bar{U}_{j\mu \nu }\right. \right.
\nonumber \\
&-&\left. \left. \frac{1}{2}\bar{B}_{i}^{a}\bar{B}_{j}^{a}V_{i\mu \nu
}V_{j\mu \nu }+\frac{1}{2}\bar{G}_{i}^{a}\bar{G}_{j}^{a}U_{i\mu \nu }U_{j\mu
\nu }\right) \right] \;,  \label{ll2}
\end{eqnarray}
fulfills all Ward identities. Moreover, this term does not
correspond to a renormalization of the parameters and fields of
$\Sigma $. This follows by noting that the counterterm (\ref{ll2})
is in fact absent in the expression (\ref{ass1}). \newline
\newline
A stable action $\widetilde{\Sigma}$ is thus obtained by adding to
the action $\Sigma $ the following expression
\begin{eqnarray}
S_{\lambda } &=&\int d^{4}x\left[ \lambda _{1}\left( \bar{B}%
_{i}^{a}B_{i}^{a}-\bar{G}_{i}^{a}G_{i}^{a}\right) \left(
\bar{V}_{j\mu \nu }V_{j\mu \nu }-\bar{U}_{j\mu \nu }U_{j\mu \nu
}\right) \right. \nonumber \\
&+&\left. \frac{\lambda^{abcd}}{16}\left( \bar{B}_{i}^{a}B_{i}^{b}-\bar{G}_{i}^{a}G_{i}^{b}%
\right)\left( \bar{B}_{j}^{c}B_{j}^{d}-\bar{G}_{j}^{c}G_{j}^{d}%
\right) \right.
\nonumber \\
&+&\left. \lambda _{3}\left( \bar{B}_{i}^{a}G_{j}^{a}V_{i\mu \nu }\bar{U}%
_{j\mu \nu }+\bar{G}_{i}^{a}G_{j}^{a}U_{i\mu \nu }\bar{U}_{j\mu \nu }+\bar{B}%
_{i}^{a}B_{j}^{a}V_{i\mu \nu }\bar{V}_{j\mu \nu }-\bar{G}%
_{i}^{a}B_{j}^{a}U_{i\mu \nu }\bar{V}_{j\mu \nu }\right. \right.   \nonumber
\\
&-&\left. \left. G_{i}^{a}B_{j}^{a}\bar{U}_{i\mu \nu }\bar{V}_{j\mu \nu }+%
\bar{G}_{i}^{a}\bar{B}_{j}^{a}U_{i\mu \nu }V_{j\mu \nu }-\frac{1}{2}%
B_{i}^{a}B_{j}^{a}\bar{V}_{i\mu \nu }\bar{V}_{j\mu \nu }+\frac{1}{2}%
G_{i}^{a}G_{j}^{a}\bar{U}_{i\mu \nu }\bar{U}_{j\mu \nu }\right. \right.
\nonumber \\
&-&\left. \left. \frac{1}{2}\bar{B}_{i}^{a}\bar{B}_{j}^{a}V_{i\mu \nu
}V_{j\mu \nu }+\frac{1}{2}\bar{G}_{i}^{a}\bar{G}_{j}^{a}U_{i\mu \nu }U_{j\mu
\nu }\right) \right] \;,  \label{l1}
\end{eqnarray}
where $\lambda _{1}, \lambda _{3}$, are free parameters, namely,
by taking as starting point the action
\begin{equation}
\widetilde{\Sigma }=S_{YM}+S_{gf}+S_{BG}+S_{aux}+S_{\lambda
}+S_{ext}\;. \label{fullaction2}
\end{equation}
The previous algebraic analysis can be repeated for the action $%
\widetilde{\Sigma }$. For the most general allowed counterterm we
find now\newpage
\begin{eqnarray}
\widetilde{\Sigma }^{c} &=&a_{0}S_{YM} +a_{1}\int d^{4}x
A_{\mu}^{a}\frac{\delta S_{YM}}{\delta {A}_{\mu
}^{a}} \nonumber \\
&+&\int {d^{4}x}\left( \left( a_{1}+a_{2}\right) \left( \Omega _{\mu }^{a}+\partial _{\mu }%
\bar{c}^{a}\right) \partial _{\mu }{c}^{a}+a_{2}gf^{abc}\left( \Omega _{\mu
}^{a}+\partial _{\mu }\bar{c}^{a}\right) A_{\mu }^{b}c^{c}-a_{2}\frac{g}{2}%
f^{abc}L^{a}c^{b}c^{c}\right)   \nonumber \\
&+&\int {d^{4}x}\left\{ \left( 2a_{3}+a_{4}\right)
\bar{B}_{i}^{a}\partial ^{2}B_{i}^{a}-\left( 2a_{3}+a_{4}\right)
\bar{G}_{i}^{a}\partial
^{2}G_{i}^{a}\right.   \nonumber \\
&-&\left. \left( a_{1}+2a_{3}+a_{4}\right)
gf^{abc}\bar{B}_{i}^{a}\left(
\partial _{\mu }{A}_{\mu }^{b}+2A_{\mu }^{b}\partial _{\mu }\right)
B_{i}^{c}+\left( 2a_{1}+2a_{3}+a_{4}\right) g^{2}f^{abd}f^{bce}\bar{B}%
_{i}^{a}A_{\mu }^{d}A_{\mu }^{e}B_{i}^{c}\right. \nonumber\\
&+&\left. \left( a_{1}+2a_{3}+a_{4}\right)
gf^{abc}\bar{G}_{i}^{a}\left(
\partial _{\mu }{A}_{\mu }^{b}+2A_{\mu }^{b}\partial _{\mu }\right)
G_{i}^{c}-\left( 2a_{1}+2a_{3}+a_{4}\right) g^{2}f^{abd}f^{bce}\bar{G}%
_{i}^{a}A_{\mu }^{d}A_{\mu }^{e}G_{i}^{c}\right.   \nonumber \\
&-&\left. a_{2}gf^{abc}c^{a}\left( \bar{Y}_{i}^{b}B_{i}^{c}+Y_{i}^{b}\bar{B}%
_{i}^{c}-\bar{X}_{i}^{b}G_{i}^{c}-X_{i}^{b}\bar{G}_{i}^{c}\right)
+\left[ \left( a_{1}+a_{3}+a_{5}\right) 2\partial _{\mu }{A}_{\nu
}^{a}\right. \right.\nonumber\\
&+&\left. \left. \left( 2a_{1}+a_{3}+a_{5}\right) gf^{abc}A_{\mu
}^{b}A_{\nu
}^{c}\right] \left( \bar{U}_{i\mu \nu }G_{i}^{a}+V_{i\mu \nu }\bar{B}%
_{i}^{a}+U_{i\mu \nu }\bar{G}_{i}^{a}-\bar{V}_{i\mu \nu }B_{i}^{a}\right)
\right.   \nonumber \\
&+&\left. \left( 4a_{3}+\widetilde{a}_{6}\right)
\frac{\lambda^{abcd}}{16}\left( \bar{B}_{i}^{a}B_{i}^{b}-\bar{G}_{i}^{a}G_{i}^{b}%
\right)\left( \bar{B}_{j}^{c}B_{j}^{d}-\bar{G}_{j}^{c}G_{j}^{d}%
\right) \right. \nonumber \\
&+&\left. \left( 2a_{3}+%
\widetilde{a}_{7}\right) \lambda _{1}\left( \bar{B}_{i}^{a}B_{i}^{a}-\bar{G}%
_{i}^{a}G_{i}^{a}\right) \left( \bar{V}_{i\mu \nu }V_{i\mu \nu }-\bar{U}%
_{i\mu \nu }U_{i\mu \nu }\right) \right.   \nonumber \\
&+&\left. \left( 2a_{3}+\widetilde{a}_{8}\right) \lambda _{3}\left( \bar{B}%
_{i}^{a}G_{j}^{a}V_{i\mu \nu }\bar{U}_{j\mu \nu }+\bar{G}%
_{i}^{a}G_{j}^{a}U_{i\mu \nu }\bar{U}_{j\mu \nu }+\bar{B}%
_{i}^{a}B_{j}^{a}V_{i\mu \nu }\bar{V}_{j\mu \nu }-\bar{G}%
_{i}^{a}B_{j}^{a}U_{i\mu \nu }\bar{V}_{j\mu \nu }\right. \right.   \nonumber
\\
&-&\left. \left. G_{i}^{a}B_{j}^{a}\bar{U}_{i\mu \nu }\bar{V}_{j\mu \nu }+%
\bar{G}_{i}^{a}\bar{B}_{j}^{a}U_{i\mu \nu }V_{j\mu \nu }-\frac{1}{2}%
B_{i}^{a}B_{j}^{a}\bar{V}_{i\mu \nu }\bar{V}_{j\mu \nu }+\frac{1}{2}%
G_{i}^{a}G_{j}^{a}\bar{U}_{i\mu \nu }\bar{U}_{j\mu \nu }\right. \right.
\nonumber \\
&-&\left. \left. \frac{1}{2}\bar{B}_{i}^{a}\bar{B}_{j}^{a}V_{i\mu
\nu }V_{j\mu \nu }+\frac{1}{2}\bar{G}_{i}^{a}\bar{G}_{j}^{a}U_{i\mu
\nu }U_{j\mu
\nu }\right) +a_{9}\zeta \left( \bar{V}_{i\mu \nu }V_{i\mu \nu }-\bar{U}%
_{i\mu \nu }U_{i\mu \nu }\right) ^{2}\right.   \nonumber \\
&+&\left. a_{10}\chi _{1}\left( \bar{V}_{i\mu \nu }\partial
^{2}V_{i\mu \nu }-\bar{U}_{i\mu \nu }\partial ^{2}U_{i\mu \nu
}\right) +a_{11}\chi _{1}\left( \bar{V}_{i\mu \nu }\partial _{\mu
}\partial _{\alpha }{V}_{i\nu \alpha }-\bar{U}_{i\mu \nu }\partial
_{\mu }\partial _{\alpha }{U}_{i\nu \alpha }\right) \right\} \;,
\label{llc}
\end{eqnarray}
As a useful check, let us show that $\widetilde{\Sigma }^{c}$ can
be reabsorbed by means of a multiplicative renormalization of the
parameters, fields and sources of $\widetilde{\Sigma }$. Setting
\begin{eqnarray}
\phi _{0} &=&Z_{\phi }^{1/2}\phi \;,  \nonumber \\
\ J_{0} &=&Z_{J}J\;,  \nonumber \\
\xi _{0} &=&Z_{\xi }\xi \;,  \label{ren0}
\end{eqnarray}
where
\begin{eqnarray}
\phi  &=&\{A,b,c,\bar{c},B,\bar{B},G,\bar{G}\}\;,  \nonumber \\
J &=&\{\Omega ,L,\bar{U},U,\bar{V},V,X,\bar{X},Y,\bar{Y}\}\;,  \nonumber \\
\xi  &=&\{g,\alpha ,\chi _{1},\chi _{2},\zeta ,\lambda _{1},
\lambda^{abcd},\lambda _{3}\}\;,  \label{ren1}
\end{eqnarray}
it follows
\begin{equation}
\widetilde{\Sigma }(\phi _{0},J_{0},\xi _{0})=\widetilde{\Sigma
}(\phi ,J,\xi )+\eta \widetilde{\Sigma }^{c}(\phi ,J,\xi
)\;+O(\eta ^{2})\;. \label{ren2}
\end{equation}
In particular, the renormalization constants are found to be
\begin{eqnarray}
\hspace{-2cm}Z_{A}^{1/2} &=&1+\eta \left( \frac{a_{0}}{2}+a_{1}\right) \;,  \\
Z_{c}^{1/2} &=&1-\eta \left( \frac{a_{1}}{2}+\frac{a_{2}}{2}\right) \;,\\
Z_{\bar{c}}^{1/2} &=&Z_{c}^{1/2}\;,  \\
Z_{b}^{1/2} &=&Z_{A}^{-1/2}\;,  \label{z1}
\end{eqnarray}
\begin{eqnarray}
Z_{\Omega } &=&Z_{c}^{1/2}\;,  \ \\
Z_{L} &=&Z_{A}\;,  \label{z2}\\
Z_{B}^{1/2}&=&Z_{\bar{B}}^{1/2}=Z_{G}^{1/2}=Z_{\bar{G}}^{1/2}=1+\eta
\left( a_{3}+\frac{a_{4}}{2}\right) \;,  \label{z4}\\
Z_{V}&=&Z_{\bar{V}}=Z_{U}=Z_{\bar{U}}=1-\eta \left( \frac{a_{0}}{2}+\frac{%
a_{4}}{2}-a_{5}\right) \;,  \label{z5}\\
Z_{X}&=&Z_{\bar{X}}=Z_{Y}=Z_{\bar{Y}}=Z_{c}^{1/2}Z_{A}^{1/2}Z_{B}^{-1/2}\;,\label{z6}\\
Z_{g} &=&1-\epsilon \frac{a_{0}}{2}\;,  \\
Z_{\alpha } &=&Z_{A}\;,  \\
Z_{\lambda _{1}} &=&1+\epsilon \left( a_{0}-2a_{5}+\widetilde{a}_{7}\right)
\;,  \\
Z_{\lambda^{abcd}} &=&1-\epsilon \left(
2a_{4}-\widetilde{a}_{6}\right) \;,
 \\
Z_{\lambda _{3}} &=&1+\epsilon \left( a_{0}-2a_{5}+\widetilde{a}_{8}\right)
\;,  \\
Z_{\chi _{1}} &=&1+\epsilon \left( a_{0}+a_{4}-2a_{5}+a_{10}\right) \;,
 \\
Z_{\chi _{2}} &=&1+\epsilon \left( a_{0}+a_{4}-2a_{5}+a_{11}\right) \;,
\\
Z_{\zeta } &=&1+\epsilon \left( 2a_{0}+2a_{4}-4a_{5}-a_{9}\right) \;.
\label{z7}
\end{eqnarray}

\subsection{Summary.}
In summary, we have been able to identify a local and polynomial
action, given in expression (\ref{fullaction2}), which displays
multiplicative renormalizability. This has been achieved by adding
to the action $\Sigma $ the term $S_{\lambda }$, eq.(\ref{l1}),
which is compatible with the complete set of Ward identities. When
the sources $\left( V_{i\mu \nu },\bar{V}_{i\mu \nu },U_{i\mu \nu
},\bar{U}_{i\mu \nu }\right)$ attain their physical value,
eq.(\ref{2q}), $S_{\lambda }$ becomes
\begin{eqnarray}
S_{\lambda }\bigg|_{\mathrm{phys}}&=&\int d^{4}x\left[ -\frac{3}{8}%
m^{2}\lambda _{1}\left( \bar{B}_{\mu \nu }^{a}B_{\mu \nu
}^{a}-\bar{G}_{\mu \nu }^{a}G_{\mu \nu }^{a}\right)+
m^{2}\frac{\lambda _{3}}{32}\left( \bar{B}_{\mu \nu }^{a}-B_{\mu
\nu }^{a}\right) ^{2} \right.\nonumber \\
&+&\left. \frac{\lambda^{abcd}}{16}\left( \bar{B}_{\mu\nu}^{a}B_{\mu\nu}^{b}-\bar{G}_{\mu\nu}^{a}G_{\mu\nu}^{b}%
\right)\left( \bar{B}_{\rho\sigma}^{c}B_{\rho\sigma}^{d}-\bar{G}_{\rho\sigma}^{c}G_{\rho\sigma}^{d}%
\right) \right] \;. \label{sla}
\end{eqnarray}
This expression reminds us of a kind of Higgs term. There are,
however, several
differences. These are due to the antisymmetric character of the fields $%
\left( B_{\mu \nu }^{a},\bar{B}_{\mu \nu }^{a},G_{\mu \nu
}^{a},\bar{G}_{\mu \nu }^{a}\right) $ with respect to the Lorentz
indices. Moreover, we remark that, while $\left( B_{\mu \nu
}^{a},\bar{B}_{\mu \nu }^{a}\right) $ are bosonic, the fields
$\left( G_{\mu \nu }^{a},\bar{G}_{\mu \nu }^{a}\right) $ are
anticommuting. With the exception of the term containing the
parameter $\lambda_3$, expression (\ref{sla}) displays thus a
supersymmetric structure, a feature supported by the fact that,
according to (\ref{bi}), the fields $\left( B_{\mu \nu
}^{a},\bar{B}_{\mu \nu }^{a},G_{\mu \nu }^{a},\bar{G}_{\mu \nu
}^{a}\right) $ transform as $BRST$ doublets. Therefore, a certain
number of cancellations among the contributions arising from these
fields might be expected in the evaluation of the Green's
functions of the model. The possible use of this supersymmetric
structure will be explored in the future, as well as its possible
consequences for the Green's functions of the model.
\newline\newline To conclude, let us give explicitly the starting
action when the sources $\left( V_{i\mu \nu },\bar{V}_{i\mu \nu
},U_{i\mu \nu },\bar{U}_{i\mu \nu }\right)$ attain their physical
value,
eq.(\ref{2q}), while the additional external sources $\left( \Omega _{\mu }^{a},L^{a},\bar{Y}%
_{i}^{a},Y_{i}^{a},\bar{X}_{i}^{a},X_{i}^{a}\right) $ are put equal
to zero.\\
\begin{eqnarray}\label{completeaction}
  S &=& S_{YM}+S_{BG}+S_{m}+S_{\lambda
  }\bigg|_{\mathrm{phys}}+S_{gf}\nonumber\\
  &=&\int d^4x\left[\frac{1}{4}F_{\mu \nu }^{a}F_{\mu \nu }^{a}+\frac{im}{4}(B-\bar{B})_{\mu\nu}^aF_{\mu\nu}^a
  +\frac{1}{4}\left( \bar{B}_{\mu \nu
}^{a}D_{\sigma }^{ab}D_{\sigma }^{bc}B_{\mu \nu }^{c}-\bar{G}_{\mu
\nu }^{a}D_{\sigma }^{ab}D_{\sigma }^{bc}G_{\mu \nu
}^{c}\right)\right.\nonumber\\
&-&\left.\frac{3}{8}%
m^{2}\lambda _{1}\left( \bar{B}_{\mu \nu }^{a}B_{\mu \nu
}^{a}-\bar{G}_{\mu \nu }^{a}G_{\mu \nu }^{a}\right)
+m^{2}\frac{\lambda _{3}}{32}\left( \bar{B}_{\mu \nu }^{a}-B_{\mu
\nu }^{a}\right) ^{2}\right.\nonumber\\&+& \left.
\frac{\lambda^{abcd}}{16}\left( \bar{B}_{\mu\nu}^{a}B_{\mu\nu}^{b}-\bar{G}_{\mu\nu}^{a}G_{\mu\nu}^{b}%
\right)\left( \bar{B}_{\rho\sigma}^{c}B_{\rho\sigma}^{d}-\bar{G}_{\rho\sigma}^{c}G_{\rho\sigma}^{d}%
\right) + \frac{\alpha }{2}b^{a}b^{a}+b^{a}\partial _{\mu }A_{\mu
}^{a}+\bar{c}^{a}\partial _{\mu }D_{\mu
}^{ab}c^{b}\vphantom{\frac{1}{4}}\right]\;. \nonumber
\\ \label{laggen}
\end{eqnarray}
Let us finally notice that each of the terms in eq.(\ref{sla}) is
invariant w.r.t. to the gauge transformations (\ref{gtm}). More
precisely, one has
\begin{eqnarray}\label{completeactioninv}
  \delta\left(S_{YM}+S_{BG}+S_{m}+S_{\lambda
  }\bigg|_{\mathrm{phys}}\right)=0\;.
\end{eqnarray}

\sect{One loop renormalization.} We now turn to the details of the
explicit one loop renormalization of the Lagrangian (\ref{laggen})
in the presence of the nonlocal operator. It is first worth noting
some of the key features of (\ref{laggen}) in relation to the
extraction of the one loop renormalization constants prior to
discussing their calculation. First, considering the case when $m$
is zero then one has a gauge theory fixed in an arbitrary linear
covariant gauge where in addition to the usual gluon and
Faddeev-Popov ghost fields there are two additional auxiliary
fields, $B^a_{\mu\nu}$ and $G^a_{\mu\nu}$ where the latter is
anticommuting. Since these fields originate in localizing the
nonlocal operator, when that operator is absent at $m$~$=$~$0$,
these new fields ought to play a completely passive role in the
(one loop) renormalization. In other words the gluon,
Faddeev-Popov ghost and quark renormalization constants ought to
be equivalent to those obtained when $B^a_{\mu\nu}$ and
$G^a_{\mu\nu}$ are formally absent. However, when they are present
the algebraic renormalization formalism has demonstrated that they
generate a new quartic interaction through (one loop)
renormalization effects\footnote{It might be useful to remark here
that, at one loop order, the invariant rank four tensor which
emerges from explicit calculations turns out to be proportional to
$g^2\left( f^{eap}f^{ebq}f^{mcp}f^{mdq} +
f^{eap}f^{ebq}f^{mdp}f^{mcq} \right)$, which fulfills in fact the
conditions (\ref{abcd}).} which is indicated by the term with the
independent coupling $\lambda^{abcd}$ in (\ref{laggen}). In other
words if one computes the $\bar{B}^a B^b \bar{B}^c B^d$ four-point
function at one loop with $\lambda^{abcd}$ initially zero, there
will be a divergent contribution at $O(g^2)$ which will be removed
by the counterterm generated by the term involving
$\lambda^{abcd}$. This is akin to the situation in $\lambda
\phi^4$ theory where the Lagrangian is multiplicatively
renormalizable in four dimensions. However, the interaction can be
replaced by a cubic vertex involving an auxiliary scalar field.
The renormalization of this version of the Lagrangian still
proceeds as usual except that the Lagrangian ceases to be
multiplicatively renormalizable since a $\phi^4$ vertex will
naturally be generated from one loop box diagrams. The standard
$\lambda \phi^4$ $\beta$-function and renormalization group
functions can still be extracted with the auxiliary field version
but one has to take account of the effects of the generation of
the extra interaction. Indeed a similar situation arises in two
dimensional four-Fermi theories where a formalism has been
developed \cite{Bondi} and used to perform three loop
calculations. The situation for our current Lagrangian is the
same. The quartic interaction is generated via loop interactions
and will be $O(g^2)$. Thus it does not need to be taken into
account for the extraction of the one loop anomalous dimensions we
are interested in. \\\\For the case when $m$ is non-zero, there is
a similar situation. The algebraic renormalization demonstrates
that the now localized mass operator $(\bar{B}^a_{\mu\nu} -
B^a_{\mu\nu}) F^{a}_{\mu\nu}$, which is dimension three, mixes
into two gauge invariant dimension two operators being those
associated with the couplings $\lambda_1$ and $\lambda_3$. In
other words computing the renormalization of the operator with a
massive gluon propagator will inevitably lead to the generation of
these two additional operators. As such this is nothing new in
that it follows the pattern already known for the renormalization
of local composite operators (See, for example, \cite{collins}).
Indeed it is reassuring that this property emerges in an elegant
way from the algebraic renormalization formalism for a {\em
localized} nonlocal operator. However, these two additional
operators do not form the complete basis of the possible dimension
two operators that higher dimensional operators can mix into when
one uses the massive theory. Since each combination of pairs of
the set $\{ B^a_{\mu\nu}, \bar{B}^a_{\mu\nu}, G^a_{\mu\nu},
\bar{G}^a_{\mu\nu} \}$ are individually gauge invariant operators,
to correctly treat the renormalization one would have to construct
the full mixing matrix for this set. Though only those
combinations with zero ghost number would be of importance. As we
are primarily focused on extracting the anomalous dimension of the
nonlocal operator itself, it will be apparent that this mixing
matrix is not immediately required and we will defer its
computation to a later article. \\\\Having outlined the status of
(\ref{laggen}) it is now evident how one goes about extracting the
renormalization constants which will lead to the anomalous
dimension of $F^a_{\mu\nu} \frac{1}{D^2} F^{a}_{\mu\nu}$. Since we
have localized this operator to $(\bar{B}^a_{\mu\nu} -
B^a_{\mu\nu}) F^{a}_{\mu\nu}$ then the anomalous dimension of
$F^a_{\mu\nu} \frac{1}{D^2} F^{a}_{\mu\nu}$ is
equivalent\footnote{Up to an overall scaling factor.} to that of
the gauge invariant operator $(\bar{B}^a_{\mu\nu} - B^a_{\mu\nu})
F^{a}_{\mu\nu}$. Therefore, we can extract the anomalous dimension
by inserting $(\bar{B}^a_{\mu\nu} - B^a_{\mu\nu}) F^{a}_{\mu\nu}$
into a $B^a_{\mu\nu} A^b_\sigma$ two-point function and compute it
using {\em massless} propagators. This is similar to how one
determines the quark mass anomalous dimension by inserting the
mass operator $\bar{\psi} \psi$ into a quark two-point function,
\cite{tarasov,larin}. For $(\bar{B}^a_{\mu\nu} - B^a_{\mu\nu})
F^{a}_{\mu\nu}$ we will need the $B^a_{\mu\nu}$ anomalous
dimension. However, we have carried out the full renormalization
of all the fields of (\ref{laggen}) at one loop by making use of
symbolic manipulation programmes. The Feynman diagrams for the
relevant Green's functions are generated with the {\sc Qgraf}
package, \cite{qgraf}, converted into {\sc Form}, \cite{form},
input notation before extracting the divergences with the {\sc
Mincer} package, \cite{mincer}. This uses dimensional
regularization in $d$~$=$~$4$~$-$~$2\epsilon$ dimensions and we
will remove the infinities with the (mass independent) $\MSbar$
renormalization scheme. If we define
\begin{equation}
\gamma_\phi(a) ~=~ \mu \frac{\partial~}{\partial\mu} \ln Z_\phi\,,
\end{equation}
for $\phi$ $\in$ $\{A^a_\mu,c^a,\psi,B^a_{\mu\nu},G^a_{\mu\nu}\}$ where
$a$~$=$~$g^2/(16\pi^2)$, then the renormalization constants give the explicit
results
\begin{eqnarray}
\gamma_A(a) &=& \left[ ( 3\alpha - 13 ) C_A + 8T_F \Nf \right] \frac{a}{6} ~+~
O(a^2) \,,\nonumber \\
\gamma_c(a) &=& ( \alpha - 3 ) C_A \frac{a}{4} ~+~ O(a^2) \,,\nonumber \\
\gamma_\psi(a) &=& \alpha C_F a ~+~ O(a^2) \,,\nonumber \\
\gamma_B(a) &=& \gamma_G(a) ~=~ ( \alpha - 3 ) C_A a ~+~ O(a^2)\,,
\end{eqnarray}
where $\Nf$ is the number of quark flavours\footnote{Although we
did not consider matter fields in the previous analysis, it turns
out that the multiplicative renormalizability of the action
$\widetilde{\Sigma }$, eq.(\ref{fullaction2}), can be extended to
the case in which spinor fields are present.}, $T^a T^a$ $=$ $C_F
I$, $f^{acd}f^{bcd}$ $=$ $C_A \delta^{ab}$ and $\mbox{tr} \left(
T^a T^b \right)$~$=$~$T_F \delta^{ab}$. For completeness we note
that the massless momentum space propagators of the fields
are\newpage
\begin{eqnarray}
\langle A^a_\mu(p) A^b_\nu(-p) \rangle &=& -~
\frac{\delta^{ab}}{p^2} \left[ \delta_{\mu\nu} ~-~ (1-\alpha)
\frac{p_\mu p_\nu}{p^2} \right] \,,\nonumber \\
\langle c^a(p) {\bar c}^b(-p) \rangle
&=& \frac{\delta^{ab}}{p^2} \,,\nonumber \\
\langle \psi(p) {\bar \psi}(-p) \rangle
&=& \frac{\pslash}{p^2} \,,\nonumber \\
\langle B^a_{\mu\nu}(p) {\bar B}^b_{\sigma\rho}(-p) \rangle &=& -~
\frac{\delta^{ab}}{2p^2} \left[ \delta_{\mu\sigma}
\delta_{\nu\rho} ~-~
\delta_{\mu\rho} \delta_{\nu\sigma} \right]\,, \nonumber \\
\langle G^a_{\mu\nu}(p) {\bar G}^b_{\sigma\rho}(-p) \rangle &=& -~
\frac{\delta^{ab}}{2p^2} \left[ \delta_{\mu\sigma}
\delta_{\nu\rho} ~-~ \delta_{\mu\rho} \delta_{\nu\sigma}
\right]\,,
\end{eqnarray}
where $p$ is the momentum. It is worth noting that the expressions
for the gluon, Faddeev-Popov ghost and quark are equivalent to
those obtained in the absence of $B^a_{\mu\nu}$ and $G^a_{\mu\nu}$
as expected. Indeed from examining the contributions from the
diagrams involving these fields it is evident that the
anticommuting property of $G^a_{\mu\nu}$ introduces the necessary
minus sign to exactly cancel the contribution from the graph
involving a $B^a_{\mu\nu}$ loop. To verify that the $B^a_{\mu\nu}$
and $G^a_{\mu\nu}$ anomalous dimensions are correct, aside from
correctly satisfying the equality demanded from the algebraic
renormalization, eq.(\ref{z4}), we have also renormalized both the
gluon-$B$ and gluon-$G$ vertices at one loop and verified that the
correct one loop $\alpha$ independent coupling constant
renormalization emerges as
\begin{equation}
\beta(a) ~=~ -~ \left[ \frac{11}{3} C_A - \frac{4}{3} T_F \Nf \right] a^2 ~+~
O(a^3) ~.
\end{equation}
Hence the renormalization of the operator $B^a_{\mu\nu}
F^{a}_{\mu\nu}$ proceeds by inserting $B^a_{\mu\nu}
F^{a}_{\mu\nu}$ into a gluon-$B$ two-point function and extracting
the divergence from the five one loop diagrams. Although we are
regarding $B^a_{\mu\nu} F^{a}_{\mu\nu}$ as multiplicatively
renormalizable, since it is of dimension three it could in
principle mix into the dimension three quark mass operator,
$\bar{\psi}\psi$. However, at one loop there are no mixed diagrams
of inserting $\bar{\psi}\psi$ into a gluon-$B$ Green's function or
of inserting $B^a_{\mu\nu} F^{a}_{\mu\nu}$ into a quark two-point
function. If we define
\begin{equation}
{\cal O}_{\mbox{\footnotesize{o}}} ~=~ Z_{\cal O} {\cal O}\,,
\end{equation}
where the subscript ${}_{\mbox{\footnotesize{o}}}$ denotes the bare object,
with
\begin{equation}
{\cal O} ~=~ B^a_{\mu\nu} F^{a}_{\mu\nu}\,,
\end{equation}
then we find
\begin{equation}
Z_{\cal O} ~=~ 1 ~+~ \left( \frac{2}{3} T_F \Nf - \frac{11}{6} C_A \right)
\frac{a}{\epsilon} ~+~ O(a^2) ~.
\end{equation}
Hence, with
\begin{equation}
 \gamma_{\cal O}(a) ~=~ ~ \mu \frac{\partial~}{\partial\mu} \ln
Z_{\cal O}\,,
\end{equation}
we deduce
\begin{equation}
\gamma_{\cal O}(a) ~=~ -~ \left[ \frac{11}{6} C_A - \frac{2}{3} T_F \Nf
\right] a ~+~ O(a^2)~.
\label{opdim}
\end{equation}
As the original operator was gauge invariant it is reassuring to
note that $\gamma_{\cal O}(a)$ is independent of $\alpha$. It is
worth underlining here that the anomalous dimension $\gamma_{\cal
O}(a)$  is equivalent to the one loop $\beta$-function, where the
overall factor of $2$ is accounted for by noting that this is
equivalent to the anomalous dimension of $m$ as opposed to that of
$m^2$. This is interesting for various reasons. First in the one
loop renormalization of two-leg higher dimension operators in
Yang-Mills theories the operators $F^a_{\mu\nu} F^{a}_{\mu\nu}$,
$D_\mu F^a_{\nu\sigma} D_\mu F^{a}_{\nu\sigma}$ and $D_\mu D_\nu
F^a_{\sigma\rho} D_\mu D_\nu F^{a}_{\sigma\rho}$ each have the
same one loop anomalous dimensions which is also the
$\beta$-function\footnote{We have checked that the dimension ten
operator $D_\mu D_\nu D_\sigma F^a_{\rho\theta} D_\mu D_\nu
D_\sigma F^{a}_{\rho\theta}$ has the same one loop anomalous
dimension too, \cite{jagunp}.}, \cite{morozov,jagclass}. What is
intriguing in the present situation is that the {\em nonlocal}
operator $F^a_{\mu\nu} \frac{1}{D^2} F^{a}_{\mu\nu}$, which has a
similar Lorentz contraction as the higher dimension operators
noted above, has an anomalous dimension which is the same at one
loop. There would appear to be no a priori reason either from the
algebraic renormalization or other methods to expect this.
Obviously, having the two loop correction to (\ref{opdim}) would
enhance our understanding of both the renormalization and
significance of this nonlocal operator. With the exception of
$F^a_{\mu\nu} F^{a}_{\mu\nu}$, the renormalization group behaviour
of the two-leg higher dimension operators is also
unknown\footnote{For details concerning the renormalization
(group) properties of $F^a_{\mu\nu} F^{a}_{\mu\nu}$ with or
without massless/massive quarks, we refer to
\cite{Tarrach:1981bi}.}. It would be interesting to pursue this
study to find out if a gauge invariant \emph{and} renormalization
group invariant mass dimension two condensate could be found using
$F^a_{\mu\nu} \frac{1}{D^2} F^{a}_{\mu\nu}$, provided the operator
condenses. Since evidence for the existence of a non-zero
dimension two condensate arises in the fitting of data for gauge
variant objects
\cite{Boucaud:2001st,Boucaud:2005rm,RuizArriola:2004en,Furui:2005bu,Boucaud:2005xn},
as a first step, it would seem natural in the light of
(\ref{opdim}) to find out whether one could extract an estimate
for the one loop renormalization group invariant condensate
$\langle \alpha_s F^a_{\mu\nu} \frac{1}{D^2} F^{a}_{\mu\nu}
\rangle$ by fitting for $1/Q^2$ power corrections in measurements
of correlations of gauge {\em invariant} operators. We refer to
\cite{Narison:2005hb,Zakharov:2005cg,Zakharov:2003nm} for a review
of the role of such $1/Q^2$ corrections which go beyond the
standard SVZ-expansion
\cite{Shifman:1978bx,Shifman:1978by,Narison:2005hb}.

\sect{Conclusions.} In this work the properties of the nonlocal
gauge invariant operator $\mathrm{Tr}\int d^{4}xF_{\mu \nu
}(D^{2})^{-1}F_{\mu \nu }$ of mass dimension 2 have been
investigated. We started by looking at the Abelian case, where
several nonlocal gauge invariant operators have been considered.
Moreover, in this case, all operators turn out to reduce to the
same expression when the classical equations of motion are
employed. All Abelian operators generalize to the non-Abelian
case. However, their
classical equivalence does not hold anymore. In particular, the operator $%
\mathrm{Tr}\int d^{4}xF_{\mu \nu }(D^{2})^{-1}F_{\mu \nu }$
exhibits differences with respect to the operator $A_{\min }^{2}$.
\newline
\newline
Albeit nonlocal, the operator $\mathrm{Tr}\int d^{4}xF_{\mu \nu
}(D^{2})^{-1}F_{\mu \nu }$ can be cast in local form by the
introduction of a suitable set of
additional fields, in contrast with the operator $A_{\min }^{2}$. A local and polynomial action has been identified, eq.(%
\ref{fullaction2}), and proven to be multiplicatively
renormalizable to all orders in the class of linear covariant
gauges by means of the algebraic renormalization. We point out
that this action possesses a finite and relatively small number of
parameters, a feature useful for higher order computations. We
have calculated the one-loop renormalization group functions of
the model. We have recovered the anomalous dimensions of the
elementary fields, if already known. In the case of the nonlocal
operator, we have found that the renormalization group behaviour
is dictated by the $\beta$-function at one-loop.
\newline
\newline
The possibility of having at our disposal a local and
renormalizable action might provide us with a consistent
framework for a future investigation of the possible existence of
the condensate $\left\langle F\frac{1}{D^{2}}F\right\rangle$.

\section*{Acknowledgements.}
The Conselho Nacional de Desenvolvimento Cient\'{i}fico e
Tecnol\'{o}gico (CNPq-Brazil), the Faperj, Funda{\c{c}}{\~{a}}o de
Amparo {\`{a}} Pesquisa do Estado do Rio de Janeiro, the SR2-UERJ
and the Coordena{\c{c}}{\~{a}}o de Aperfei{\c{c}}oamento de
Pessoal de N{\'\i}vel Superior (CAPES) are gratefully acknowledged
for financial support. \newline \noindent D.~Dudal is a
postdoctoral fellow of the \emph{Special Research Fund} of Ghent
University.\newline \noindent R.~F.~Sobreiro would like to thank
the warm hospitality of the Department of Mathematical Physics and
Astronomy of Ghent University.

\appendix
\sect{Properties of the functional $f_{A}[u]$.} \label{apb}In this
 Appendix we recall some useful properties of the functional
$f_{A}[u]$
\begin{equation}
f_{A}[u]\equiv \mathrm{Tr}\int d^{4}x\,A_{\mu }^{u}A_{\mu
}^{u}=\mathrm{Tr}\int d^{4}x\left( u^{\dagger }A_{\mu
}u+\frac{i}{g}u^{\dagger }\partial _{\mu }u\right) \left(
u^{\dagger }A_{\mu }u+\frac{i}{g}u^{\dagger }\partial _{\mu
}u\right) \;. \label{fa}
\end{equation}
For a given gauge field configuration $A_{\mu }$, $f_{A}[u]$ is a functional
defined on the gauge orbit of $A_{\mu }$. Let $\mathcal{A}$ be the space of
connections $A_{\mu }^{a}$ with finite Hilbert norm $||A||$, \textit{i.e.}
\begin{equation}
||A||^{2}=\mathrm{Tr}\int d^{4}x\,A_{\mu }A{_{\mu }=}\frac{1}{2}\int
d^{4}xA_{\mu }^{a}A_{\mu }^{a}<+\infty \;, \label{norm0}
\end{equation}
and let $\mathcal{U}$ be the space of local gauge transformations $u$ such
that the Hilbert norm $||u^{\dagger }\partial {u}||$ is finite too, namely
\begin{equation}
||u^{\dagger }\partial {u}||^{2}=\mathrm{Tr}\int d^{4}x\,\left(
u^{\dagger }\partial _{\mu }u\right) \left( u^{\dagger }\partial
_{\mu }u\right) <+\infty \;. \label{norm1}
\end{equation}
\noindent The following proposition holds
\cite{Semenov,Zwanziger:1990tn,Dell'Antonio:1989jn,Dell'Antonio:1991xt,vanBaal:1991zw}
\begin{itemize}
\item  \underline{Proposition}\newline
The functional $f_{A}[u]$ achieves its absolute minimum on the gauge orbit
of $A_{\mu }$.
\end{itemize}
\noindent This proposition means that there exists a $h\in
\mathcal{U}$ such that
\begin{eqnarray}
\delta f_{A}[h] &=&0\;,  \label{impl0} \\
\delta ^{2}f_{A}[h] &\ge &0\;,  \label{impl1} \\
f_{A}[h] &\le &f_{A}[u]\;,\;\;\;\;\;\;\;\forall \,u\in
\mathcal{U}\;. \label{impl2}
\end{eqnarray}
The operator $A_{\min }^{2}$ is thus given by
\begin{equation}
A_{\min }^{2}=\min_{\left\{ u\right\} }\mathrm{Tr}\int
d^{4}x\,A_{\mu }^{u}A_{\mu }^{u}=f_{A}[h]\;.  \label{a2min}
\end{equation}
Let us give a look at the two conditions (\ref{impl0}) and
(\ref{impl1}). To evaluate $\delta f_{A}[h]$ and $\delta
^{2}f_{A}[h]$ we set\footnote{The case of the gauge group $SU(N)$
is considered here.}
\begin{equation}
v=he^{ig\omega }=he^{ig\omega ^{a}T^{a}}\;,  \label{set0}
\end{equation}
\begin{equation}
\left[ T^{a},T^{b}\right] =if^{abc}\;,\;\;\;\;\;\mathrm{Tr}\left( T^{a}T^{b}\right) =%
\frac{1}{2}\delta ^{ab}\;,  \label{st000}
\end{equation}
where $\omega $ is an infinitesimal hermitian matrix and we
compute the linear and quadratic terms of the expansion of the
functional $f_{A}[v]$ in power series of $\omega $. Let us first
obtain an expression for $A_{\mu }^{v}$
\begin{eqnarray}
A_{\mu }^{v} &=&v^{\dagger }A_{\mu }v+\frac{i}{g}v^{\dagger }\partial _{\mu
}v  \nonumber \\
&=&e^{-ig\omega }h^{\dagger }A_{\mu }he^{ig\omega }+\frac{i}{g}e^{-ig\omega
}\left( h^{\dagger }\partial _{\mu }h\right) e^{ig\omega }+\frac{i}{g}%
e^{-ig\omega }\partial _{\mu }e^{ig\omega }  \nonumber \\
&=&e^{-ig\omega }A_{\mu }^{h}e^{ig\omega }+\frac{i}{g}e^{-ig\omega }\partial
_{\mu }e^{ig\omega }\;.  \label{orbit0}
\end{eqnarray}
Expanding up to the order $\omega ^{2}$, we get
\begin{eqnarray}
A_{\mu }^{v} &=&\left( 1-ig\omega -g^{2}\frac{\omega ^{2}}{2}\right) A_{\mu
}^{h}\left( 1+ig\omega -g^{2}\frac{\omega ^{2}}{2}\right) +\frac{i}{g}\left(
1-ig\omega -g^{2}\frac{\omega ^{2}}{2}\right) \partial _{\mu }\left(
1+ig\omega -g^{2}\frac{\omega ^{2}}{2}\right)  \nonumber \\
&=&\left( 1-ig\omega -g^{2}\frac{\omega ^{2}}{2}\right) \left( A_{\mu
}^{h}+igA_{\mu }^{h}\omega -g^{2}A_{\mu }^{h}\frac{\omega ^{2}}{2}\right) +
\nonumber \\
&+&\frac{i}{g}\left( 1-ig\omega -g^{2}\frac{\omega ^{2}}{2}\right) \left(
ig\partial _{\mu }\omega -\frac{g^{2}}{2}\left( \partial _{\mu }\omega
\right) \omega -\frac{g^{2}}{2}\omega \left( \partial _{\mu }\omega \right)
\right)  \nonumber \\
&=&A_{\mu }^{h}+igA_{\mu }^{h}\omega -\frac{g^{2}}{2}A_{\mu }^{h}\omega
^{2}-ig\omega A_{\mu }^{h}+g^{2}\omega A_{\mu }^{h}\omega -\frac{g^{2}}{2}%
\omega ^{2}A_{\mu }^{h}  \nonumber \\
&+&\frac{i}{g}\left( ig\partial _{\mu }\omega -\frac{g^{2}}{2}\left(
\partial _{\mu }\omega \right) \omega -\frac{g^{2}}{2}\omega \partial _{\mu
}\omega +g^{2}\omega \partial _{\mu }\omega \right) +O(\omega ^{3})\;,
\label{ex1}
\end{eqnarray}
from which it follows
\begin{equation}
A_{\mu }^{v}=A_{\mu }^{h}+ig[A_{\mu }^{h},\omega ]+\frac{g^{2}}{2}[[\omega
,A_{\mu }^{h}],\omega ]-\partial _{\mu }\omega +i\frac{g}{2}[\omega
,\partial _{\mu }\omega ]+O(\omega ^{3})\;,  \label{A0}
\end{equation}
We now evaluate
\begin{eqnarray}
f_{A}[v] &=&\mathrm{Tr}\int d^{4}xA_{\mu }^{u}A_{\mu }^{u}  \nonumber \\
&=&\mathrm{Tr}\int d^{4}x\,\left[ \left( A_{\mu }^{h}+ig[A_{\mu }^{h},\omega ]+\frac{%
g^{2}}{2}[[\omega ,A_{\mu }^{h}],\omega ]-\partial _{\mu }\omega +i\frac{g}{2%
}[\omega ,\partial _{\mu }\omega ]+O(\omega ^{3})\right) \times
\right.
\nonumber \\
& &\left. \left( A_{\mu }^{h}+ig[A_{\mu }^{h},\omega ]+\frac{g^{2}}{2}[%
[\omega ,A_{\mu }^{h}],\omega ]-\partial _{\mu }\omega +i\frac{g}{2}[\omega
,\partial _{\mu }\omega ]+O(\omega ^{3})\right) \right]  \nonumber \\
&=&\mathrm{Tr}\int d^{4}x\,\left\{ A_{\mu }^{h}A_{\mu
}^{h}+igA_{\mu }^{h}[A_{\mu
}^{h},\omega ]+g^{2}A_{\mu }^{h}\omega {A}_{\mu }^{h}\omega -\frac{g^{2}}{2}%
A_{\mu }^{h}A_{\mu }^{h}\omega ^{2}-\frac{g^{2}}{2}A_{\mu }^{h}\omega
^{2}A_{\mu }^{h}-A_{\mu }^{h}\partial _{\mu }\omega \right.  \nonumber \\
&+&\left. i\frac{g}{2}A_{\mu }^{h}[\omega ,\partial _{\mu }\omega
]+ig[A_{\mu }^{h},\omega ]A_{\mu }^{h}-g^{2}[A_{\mu }^{h},\omega
][A_{\mu }^{h},\omega ]-ig[A_{\mu }^{h},\omega ]\partial _{\mu
}\omega +g^{2}\omega
A_{\mu }^{h}\omega A_{\mu }^{h}\right.  \nonumber \\
&-&\left. \frac{g^{2}}{2}A_{\mu }^{h}\omega ^{2}A_{\mu }^{h}-\frac{g^{2}}{2}%
\omega ^{2}A_{\mu }^{h}A_{\mu }^{h}-\partial _{\mu }\omega A_{\mu
}^{h}-ig\partial _{\mu }\omega [A_{\mu }^{h},\omega ]+\partial _{\mu }\omega
\partial _{\mu }\omega +i\frac{g}{2}[\omega ,\partial _{\mu }\omega ]A_{\mu
}^{h}\right\} +O(\omega ^{3})\nonumber\\
 &=&f_{A}[h]-\mathrm{Tr}\int d^{4}x\,\left\{
A_{\mu }^{h},\partial _{\mu }\omega
\right\} +\mathrm{Tr}\int d^{4}x\,\left( g^{2}A_{\mu }^{h}\omega A_{\mu }^{h}\omega -%
\frac{g^{2}}{2}A_{\mu }^{h}A_{\mu }^{h}\omega ^{2}-\frac{g^{2}}{2}A_{\mu
}^{h}\omega ^{2}A_{\mu }^{h}\right.  \nonumber \\
&-&\left. g^{2}[A_{\mu }^{h},\omega ][A_{\mu }^{h},\omega
]+g^{2}\omega A_{\mu }^{h}\omega A_{\mu
}^{h}-\frac{g^{2}}{2}A_{\mu }^{h}\omega ^{2}A_{\mu
}^{h}-\frac{g^{2}}{2}\omega ^{2}A_{\mu }^{h}A_{\mu }^{h}\right)
+\mathrm{Tr}\int d^{4}x\,\left( \partial _{\mu }\omega \partial
_{\mu }\omega \right.
\nonumber \\
&+&\left. i\frac{g}{2}[\omega ,\partial _{\mu }\omega ]A_{\mu
}^{h}-ig\partial _{\mu }\omega [A_{\mu }^{h},\omega ]-ig[A_{\mu
}^{h},\omega ]\partial _{\mu }\omega +i\frac{g}{2}A_{\mu
}^{h}[\omega ,\partial _{\mu }\omega ]\right) +O(\omega ^{3})
\nonumber
\end{eqnarray}
\begin{eqnarray}
&=&f_{A}[h]+2\int {d^{4}x}\,tr\left( \omega \partial _{\mu
}{A}_{\mu }^{h}\right) +\int {d^{4}x}\,tr\left\{ 2g^{2}\omega
{A}_{\mu }^{h}\omega
A_{\mu }^{h}-2g^{2}A_{\mu }^{h}A_{\mu }^{h}\omega ^{2}\right.  \nonumber \\
&-&\left. g^{2}\left( A_{\mu }^{h}\omega -\omega {A}_{\mu }^{h}\right)
\left( A_{\mu }^{h}\omega -\omega {A}_{\mu }^{h}\right) \right\} +\int {%
d^{4}x}\,tr\left( \partial _{\mu }\omega \partial _{\mu }\omega +i\frac{g}{2}%
\omega \partial _{\mu }\omega {A}_{\mu }^{h}-i\frac{g}{2}\partial _{\mu
}\omega \omega {A}_{\mu }^{h}\right.  \nonumber \\
&-&\left. ig\partial _{\mu }\omega {A}_{\mu }^{h}\omega +ig\partial _{\mu
}\omega \omega {A}_{\mu }^{h}-igA_{\mu }^{h}\omega \partial _{\mu }\omega
+ig\omega {A}_{\mu }^{h}\partial _{\mu }\omega +i\frac{g}{2}A_{\mu
}^{h}\omega \partial _{\mu }\omega -i\frac{g}{2}A_{\mu }^{h}\partial _{\mu
}\omega \omega \right) +O(\omega ^{3})  \nonumber \\
&=&f_{A}[h]+2\mathrm{Tr}\int d^{4}x\left( \,\omega \partial _{\mu
}A_{\mu }^{h}\right) +\mathrm{Tr}\int d^{4}x\,\left( \partial
_{\mu }\omega
\partial _{\mu }\omega +ig\omega \partial _{\mu }\omega {A}_{\mu
}^{h}-ig\partial _{\mu
}\omega \omega {A}_{\mu }^{h}\right.  \nonumber \\
&-&\left. 2ig\partial _{\mu }\omega A_{\mu }^{h}\omega +2ig\partial
_{\mu }\omega \omega A_{\mu }^{h}\right) +O(\omega ^{3})\;.
\end{eqnarray}
Thus
\begin{eqnarray}
f_{A}[v] &=&f_{A}[h]+2\mathrm{Tr}\int d^{4}x\,\left( \omega
\partial _{\mu }A_{\mu }^{h}\right) +\mathrm{Tr}\int d^{4}x\,\left(
\partial _{\mu }\omega \partial _{\mu }\omega +ig\omega \partial
_{\mu }\omega A_{\mu }^{h}-ig\partial _{\mu
}\omega \omega A_{\mu }^{h}\right.  \nonumber \\
&-&\left. ig\left( \partial _{\mu }\omega \right) A_{\mu }^{h}\omega
+ig\left( \partial _{\mu }\omega \right) \omega A_{\mu }^{h}\right)
+O(\omega ^{3})  \nonumber \\
&=&f_{A}[h]+2\mathrm{Tr}\int d^{4}x\,\left( \omega \partial _{\mu
}A_{\mu }^{h}\right) +\mathrm{Tr}\int d^{4}x\,\left\{ \partial
_{\mu }\omega \left(
\partial _{\mu }\omega -ig\left[ A_{\mu }^{h},\omega \right]
\right) \right\} +O(\omega ^{3})\;. \nonumber  \\ \label{f1}
\end{eqnarray}
Finally
\begin{equation}
f_{A}[v]=f_{A}[h]+2\mathrm{Tr}\int d^{4}x\,\left( \omega \partial
_{\mu }A_{\mu }^{h}\right) -\mathrm{Tr}\int d^{4}x\,\omega
\partial _{\mu }D_{\mu }(A^{h})\omega +O(\omega ^{3})\;,
\label{func2}
\end{equation}
so that
\begin{eqnarray}
\delta f_{A}[h] &=&0\;\;\;\Rightarrow \;\;\;\partial _{\mu }A_{\mu
}^{h}\;=\;0\;,  \nonumber \\
\delta ^{2}f_{A}[h] &>&0\;\;\;\Rightarrow \;\;\;-\partial _{\mu }D{_{\mu }(}%
A^{h}{)}\;>\;0\;.  \label{func3}
\end{eqnarray}
We see therefore that the set of field configurations fulfilling conditions (%
\ref{func3}), \textit{i.e.} defining relative minima of the functional $%
f_{A}[u]$, belong to the so called Gribov region $\Omega $, which is defined
as
\begin{equation}
\Omega =\left.\{A_{\mu }\right|\partial _{\mu }A_{\mu
}=0\;\mathrm{and}\;-\partial _{\mu }D_{\mu }(A)>0\}\;.
\label{gribov0}
\end{equation}
Let us proceed now by showing that the transversality condition,
$\partial
_{\mu }A_{\mu }^{h}=0$, can be solved for $h=h(A)$ as a power series in $%
A_{\mu }$. We start from
\begin{equation}
A_{\mu }^{h}=h^{\dagger }A_{\mu }h+\frac{i}{g}h^{\dagger }\partial _{\mu
}h\;,  \label{Ah0}
\end{equation}
with
\begin{equation}
h=e^{ig\phi }=e^{ig\phi ^{a}T^{a}}\;.  \label{h0}
\end{equation}
Let us expand $h$ in powers of $\phi $
\begin{equation}
h=1+ig\phi -\frac{g^{2}}{2}\phi ^{2}+O(\phi ^{3})\;.  \label{hh1}
\end{equation}
From equation (\ref{Ah0}) we have
\begin{equation}
A_{\mu }^{h}=A_{\mu }+ig[A_{\mu },\phi ]+g^{2}\phi A_{\mu }\phi -\frac{g^{2}%
}{2}A_{\mu }\phi ^{2}-\frac{g^{2}}{2}\phi ^{2}A_{\mu }-\partial _{\mu }\phi
+i\frac{g}{2}[\phi ,\partial _{\mu }]+O(\phi ^{3})\;.  \label{A1}
\end{equation}
Thus, condition $\partial _{\mu }A_{\mu }^{h}=0$, gives
\begin{eqnarray}
\partial ^{2}\phi &=&\partial _{\mu }A+ig[\partial _{\mu }A_{\mu },\phi
]+ig[A_{\mu },\partial _{\mu }\phi ]+g^{2}\partial _{\mu }\phi A_{\mu }\phi
+g^{2}\phi \partial _{\mu }A_{\mu }\phi +g^{2}\phi A_{\mu }\partial _{\mu
}\phi   \nonumber \\
&-&\frac{g^{2}}{2}\partial _{\mu }A_{\mu }\phi ^{2}-\frac{g^{2}}{2}A_{\mu
}\partial _{\mu }\phi \phi -\frac{g^{2}}{2}A_{\mu }\phi \partial _{\mu }\phi
-\frac{g^{2}}{2}\partial _{\mu }\phi \phi A_{\mu }-\frac{g^{2}}{2}\phi
\partial _{\mu }\phi A_{\mu }-\frac{g^{2}}{2}\phi ^{2}\partial _{\mu }A_{\mu
}  \nonumber \\
&+&i\frac{g}{2}[\phi ,\partial ^{2}\phi ]+O(\phi ^{3})\;.  \label{hh2}
\end{eqnarray}
This equation can be solved iteratively for $\phi $ as a power series in $%
A_{\mu }$, namely
\begin{equation}
\phi =\frac{1}{\partial ^{2}}\partial _{\mu }A_{\mu }+i\frac{g}{\partial ^{2}%
}\left[ \partial A,\frac{\partial A}{\partial ^{2}}\right] +i\frac{g}{%
\partial ^{2}}\left[ A_{\mu },\partial _{\mu }\frac{\partial A}{\partial ^{2}%
}\right] +\frac{i}{2}\frac{g}{\partial ^{2}}\left[ \frac{\partial A}{%
\partial ^{2}},\partial A\right] +O(A^{3})\;,  \label{phi0}
\end{equation}
so that
\begin{eqnarray}
A_{\mu }^{h} &=&A_{\mu }-\frac{1}{\partial ^{2}}\partial _{\mu }\partial A-ig%
\frac{\partial _{\mu }}{\partial ^{2}}\left[ A_{\nu },\partial _{\nu }\frac{%
\partial A}{\partial ^{2}}\right] -i\frac{g}{2}\frac{\partial _{\mu }}{%
\partial ^{2}}\left[ \partial A,\frac{1}{\partial ^{2}}\partial A\right]
\nonumber \\
&+&ig\left[ A_{\mu },\frac{1}{\partial ^{2}}\partial A\right] +i\frac{g}{2}%
\left[ \frac{1}{\partial ^{2}}\partial A,\frac{\partial _{\mu }}{\partial
^{2}}\partial A\right] +O(A^{3})\;.  \label{minn2}
\end{eqnarray}
Expression (\ref{minn2}) can be written in a more useful way,
given in eq.(\ref{min0}). In fact
\begin{eqnarray}
A_{\mu }^{h} &=&\left( \delta _{\mu \nu }-\frac{\partial _{\mu }\partial
_{\nu }}{\partial ^{2}}\right) \left( A_{\nu }-ig\left[ \frac{1}{\partial
^{2}}\partial A,A_{\nu }\right] +\frac{ig}{2}\left[ \frac{1}{\partial ^{2}}%
\partial A,\partial _{\nu }\frac{1}{\partial ^{2}}\partial A\right] \right)
+O(A^{3})  \nonumber \\
&=&A_{\mu }-ig\left[ \frac{1}{\partial ^{2}}\partial A,A_{\mu }\right] +%
\frac{ig}{2}\left[ \frac{1}{\partial ^{2}}\partial A,\partial _{\mu }\frac{1%
}{\partial ^{2}}\partial A\right] -\frac{\partial _{\mu }}{\partial ^{2}}%
\partial A+ig\frac{\partial _{\mu }}{\partial ^{2}}\partial _{\nu }\left[
\frac{1}{\partial ^{2}}\partial A,A_{\nu }\right]   \nonumber \\
&-&i\frac{g}{2}\frac{\partial _{\mu }}{\partial ^{2}}\partial _{\nu }\left[
\frac{\partial A}{\partial ^{2}},\frac{\partial _{\nu }}{\partial ^{2}}%
\partial A\right] +O(A^{3})  \nonumber \\
&=&A_{\mu }-\frac{\partial _{\mu }}{\partial ^{2}}\partial A+ig\left[ A_{\mu
},\frac{1}{\partial ^{2}}\partial A\right] +\frac{ig}{2}\left[ \frac{1}{%
\partial ^{2}}\partial A,\partial _{\mu }\frac{1}{\partial ^{2}}\partial
A\right] +ig\frac{\partial _{\mu }}{\partial ^{2}}\left[ \frac{\partial
_{\nu }}{\partial ^{2}}\partial A,A_{\nu }\right]   \nonumber \\
&+&i\frac{g}{2}\frac{\partial _{\mu }}{\partial ^{2}}\left[ \frac{\partial A%
}{\partial ^{2}},\partial A\right] +O(A^{3})  \label{hhh3}
\end{eqnarray}
which is precisely expression (\ref{minn2}). The transverse field
given in eq.(\ref {min0}) enjoys the property of being gauge
invariant order by order in the
coupling constant $g$. Let us work out the transformation properties of $%
\phi _{\nu }$ under a gauge transformation
\begin{equation}
\delta A_{\mu }=-\partial _{\mu }\omega +ig[A_{\mu },\omega ]\;.
\label{gauge3}
\end{equation}
We have, up to the order $O(g^{2})$,
\begin{eqnarray}
\delta \phi _{\nu } &=&-\partial _{\nu }\omega +ig\left[ \frac{1}{\partial
^{2}}\partial A,\partial _{\nu }\omega \right] -i\frac{g}{2}\left[ \omega
,\partial _{\nu }\frac{1}{\partial ^{2}}\partial A\right] -i\frac{g}{2}%
\left[ \frac{\partial A}{\partial ^{2}},\partial _{\nu }\omega \right]
+O(g^{2})  \nonumber \\
&=&-\partial _{\nu }\omega +i\frac{g}{2}\left[ \frac{1}{\partial ^{2}}%
\partial A,\partial _{\nu }\omega \right] +i\frac{g}{2}\left[ \partial _{\nu
}\frac{1}{\partial ^{2}}\partial A,\omega \right] +O(g^{2})\;.  \label{gg2}
\end{eqnarray}
Therefore
\begin{equation}
\delta \phi _{\nu }=-\partial _{\nu }\left( \omega -i\frac{g}{2}\left[ \frac{%
\partial A}{\partial ^{2}},\omega \right] \right) +O(g^{2})\;,  \label{phi1}
\end{equation}
from which the gauge invariance of $A_{\mu }^{h}$ is established.\newline
\newline
Finally, let us work out the expression of $A_{\mathrm{min}}^{2}$ as a power
series in $A_{\mu }$.
\begin{eqnarray}
A_{\mathrm{min}}^{2} &=&\mathrm{Tr}\int d^{4}x\,A_{\mu }^{h}A_{\mu }^{h}  \nonumber \\
&=&\mathrm{Tr}\int d^{4}x\,\left[ \phi _{\mu }\left( \delta _{\mu \nu }-\frac{%
\partial _{\mu }\partial _{\nu }}{\partial ^{2}}\right) \phi _{\nu }\right]
\nonumber \\
&=&\mathrm{Tr}\int d^{4}x\,\left[ \left( A_{\mu }-ig\left[ \frac{1}{\partial ^{2}}%
\partial A,A_{\mu }\right] +\frac{ig}{2}\left[ \frac{1}{\partial ^{2}}%
\partial A,\partial _{\mu }\frac{1}{\partial ^{2}}\partial A\right] \right)
\times \right.  \nonumber \\
&&\left.  \left( \delta _{\mu \nu }-\frac{\partial _{\mu }\partial
_{\nu }}{\partial ^{2}}\right) \left( A_{\nu }-ig\left[
\frac{1}{\partial
^{2}}\partial A,A_{\nu }\right] +\frac{ig}{2}\left[ \frac{1}{\partial ^{2}}%
\partial A,\partial _{\nu }\frac{1}{\partial ^{2}}\partial A\right] \right)
\right]  \nonumber \\
&=&\mathrm{Tr}\int d^{4}x\,\left\{ A_{\mu }\left( \delta _{\mu \nu
}-\frac{\partial _{\mu }\partial _{\nu }}{\partial ^{2}}\right)
A_{\nu }-2ig\left( A_{\nu
}-\partial _{\nu }\frac{\partial A}{\partial ^{2}}\right) \left[ \frac{%
\partial A}{\partial ^{2}},A_{\nu }\right] \right.  \nonumber \\
&&\left. +ig\left( A_{\nu }-\partial _{\nu }\frac{\partial A}{\partial ^{2}}%
\right) \left[ \frac{\partial A}{\partial ^{2}},\partial _{\nu }\frac{%
\partial A}{\partial ^{2}}\right] \right\} +O(A^{4})  \nonumber \\
&=&\mathrm{Tr}\int d^{4}x\,\left\{ A_{\mu }\left( \delta _{\mu \nu
}-\frac{\partial _{\mu }\partial _{\nu }}{\partial ^{2}}\right)
A_{\nu }-2igA_{\nu }\left[ \frac{\partial A}{\partial ^{2}},A_{\nu
}\right] +2ig\frac{\partial _{\nu }\partial A}{\partial
^{2}}\left[ \frac{\partial A}{\partial ^{2}},A_{\nu
}\right] \right.  \nonumber \\
&&\left. +igA_{\nu }\left[ \frac{\partial A}{\partial ^{2}},\partial _{\nu }%
\frac{\partial A}{\partial ^{2}}\right] -ig\frac{\partial _{\nu }\partial A}{%
\partial ^{2}}\left[ \frac{\partial A}{\partial ^{2}},\partial _{\nu }\frac{%
\partial A}{\partial ^{2}}\right] \right\} +O(A^{4})  \nonumber \\
&=&\frac{1}{2}\int d^{4}x\left[ A_{\mu }^{a}\left( \delta _{\mu \nu }-\frac{%
\partial _{\mu }\partial _{\nu }}{\partial ^{2}}\right) A_{\nu
}^{a}-2gf^{abc}\frac{\partial _{\nu }\partial A^{a}}{\partial ^{2}}\frac{%
\partial A^{b}}{\partial ^{2}}A_{\nu }^{c}-gf^{abc}A_{\nu }^{a}\frac{%
\partial A^{b}}{\partial ^{2}}\frac{\partial _{\nu }\partial A^{c}}{\partial
^{2}}\right] +O(A^{4})\;.  \nonumber \\
&&  \label{a2em}
\end{eqnarray}
leading to the result quoted in eq.(\ref{min1}).\newline\newline
We conclude this Appendix by noting that, due to gauge invariance, $A_{%
\mathrm{min}}^{2}$ can be rewritten in a manifestly invariant way
in terms of $F_{\mu \nu }$ and the covariant derivative $D_{\mu }$
\cite{Zwanziger:1990tn}, see eq.(\ref{zzw}).

\sect{Properties of the Stueckelberg term.} \label{apc} In this
Appendix we derive some useful properties of the non-Abelian
Stueckelberg term  $\mathcal{O}_{S}$ \cite{Ruegg:2003ps}, defined
by the equations (\ref{stueck0})-(\ref{U0}). The expression
(\ref{stueck0}) is left invariant by the gauge transformations
given in eq.(\ref{gauge4}). In fact
\begin{eqnarray}
\left( A_{\mu }-\frac{i}{g}U^{-1}\partial _{\mu }U\right)
&\rightarrow &V^{-1}\left( A_{\mu }-\frac{i}{g}U^{-1}\partial _{\mu
}U\right) V\;. \label{stt1}
\end{eqnarray}
Thus
\begin{eqnarray}
\mathrm{Tr}\left( A_{\mu }-\frac{i}{g}U^{-1}\partial _{\mu }U\right)
^{2} &\rightarrow &\mathrm{Tr}\left( A_{\mu
}-\frac{i}{g}U^{-1}\partial _{\mu }U\right) ^{2}\;. \label{stt2}
\end{eqnarray}
Let us look now at the equations of motion of the Stueckelberg
field $\phi ^{a}$, as expressed in eq.(\ref{eq2}), from which
\begin{equation}
\partial _{\mu }A_{\mu }-\frac{i}{g}\partial _{\mu }\left( U^{-1}\partial
_{\mu }U\right) -\left[ A_{\mu },U^{-1}\partial _{\mu }U\right] =0\;.
\label{eq3}
\end{equation}
Expanding the term $U^{-1}\partial _{\mu }U$ in power series of $\phi ^{a}$
\begin{eqnarray}
U^{-1}\partial _{\mu }U &=&e^{-ig\phi ^{a}T^{a}}\partial _{\mu }e^{ig\phi
^{a}T^{a}}  \nonumber \\
&=&\left( 1-ig\phi ^{a}T^{a}-\frac{g^{2}}{2}\phi ^{a}T^{a}\phi
^{b}T^{b}\right) \partial _{\mu }\left( 1+ig\phi ^{a}T^{a}-\frac{g^{2}}{2}%
\phi ^{a}T^{a}\phi ^{b}T^{b}\right) +O(\phi ^{3})  \nonumber \\
&=&\left( 1-ig\phi ^{a}T^{a}-\frac{g^{2}}{2}\phi ^{a}T^{a}\phi
^{b}T^{b}\right) \left( ig\partial _{\mu }\phi ^{a}T^{a}-\frac{g^{2}}{2}%
\partial _{\mu }\phi ^{a}\phi ^{b}T^{a}T^{b}-\frac{g^{2}}{2}\phi
^{a}\partial _{\mu }\phi ^{b}T^{a}T^{b}\right)  \nonumber \\
&=&igT^{a}\partial _{\mu }\phi ^{a}-\frac{g^{2}}{2}\left( \partial _{\mu
}\phi ^{a}\right) \phi ^{b}T^{a}T^{b}-\frac{g^{2}}{2}\phi ^{a}\partial _{\mu
}\phi ^{b}T^{a}T^{b}+g^{2}\phi ^{a}\partial _{\mu }\phi
^{b}T^{a}T^{b}+O(\phi ^{3})  \nonumber \\
&=&igT^{a}\partial _{\mu }\phi ^{a}-\frac{g^{2}}{2}\left( \partial _{\mu
}\phi ^{a}\right) \phi ^{b}[T^{a},T^{b}]+O(\phi ^{3})\;,  \label{stkm1}
\end{eqnarray}
yielding
\begin{equation}
U^{-1}\partial _{\mu }U=igT^{a}\partial _{\mu }\phi ^{a}-\frac{g^{2}}{2}%
iT^{c}f^{abc}\left( \partial _{\mu }\phi ^{a}\right) \phi ^{b}+O(\phi
^{3})\;.  \label{stkm2}
\end{equation}
After substitution of expression (\ref{stkm2}) in eq.(\ref{eq3}), we have
\begin{eqnarray}
0 &=&\partial _{\mu }A_{\mu }^{a}T^{a}+\partial _{\mu }\left( T^{a}\partial
_{\mu }\phi ^{a}-\frac{g}{2}T^{c}f^{abc}\partial _{\mu }\phi ^{a}\phi
^{b}\right) -igA_{\mu }^{b}\partial _{\mu }\phi ^{c}[T^{b},T^{c}]+\mathrm{%
higher\;order\;terms}\;\;  \nonumber \\
&=&T^{a}\left( \partial A^{a}+\partial ^{2}\phi ^{a}-\frac{g}{2}%
f^{abc}\left( \partial ^{2}\phi ^{b}\right) \phi ^{c}+gf^{abc}A_{\mu
}^{b}\partial _{\mu }\phi ^{c}\right) +\mathrm{higher\;order\;terms}\;,
\label{stkm3}
\end{eqnarray}
from which
\begin{equation}
\partial ^{2}\phi ^{a}=-\partial A^{a}-gf^{abc}A_{\mu }^{b}\partial _{\mu
}\phi ^{c}+\frac{g}{2}f^{abc}\left( \partial ^{2}\phi ^{b}\right) \phi ^{c}+%
\mathrm{higher\;order\;terms}\;,  \label{stkm4}
\end{equation}
and
\begin{equation}
\phi ^{a}=-\frac{1}{\partial ^{2}}\partial A^{a}-\frac{g}{\partial ^{2}}%
\left( f^{abc}A_{\mu }^{b}\partial _{\mu }\phi ^{c}-\frac{g}{2}f^{abc}\left(
\partial ^{2}\phi ^{b}\right) \phi ^{c}\right) +\mathrm{higher\;order\;terms}%
\;.  \label{stkm5}
\end{equation}
Finally, substituting recursively for $\phi$, we obtain the
expression (\ref{stueck1}).

\end{document}